\documentclass[journal=jpcbfk,manuscript=article]{achemso}

\usepackage{amssymb}
\usepackage{graphicx}
\usepackage{hyperref}
\usepackage{nameref}
\usepackage{amsmath}
\usepackage{siunitx}
\usepackage{float}
\usepackage{multirow}
\usepackage{enumitem}
\usepackage[utf8]{inputenc}
\usepackage[version=4]{mhchem}

\title{Local ion environment in polyamide membranes revealed by molecular dynamics}

\author{Nathanael S. Schwindt}
\affiliation[CU]{Department of Chemical \& Biological Engineering, University of Colorado Boulder, Boulder, CO, USA 80309}
\author{Anthony P. Straub}
\affiliation[CU]{Department of Civil, Environmental, \& Architectural Engineering, University of Colorado Boulder, Boulder, CO, USA 80309}
\author{Michael F. Toney}
\affiliation[CU]{Department of Chemical \& Biological Engineering, University of Colorado Boulder, Boulder, CO, USA 80309}
\author{Michael R. Shirts}
\affiliation[CU]{Department of Chemical \& Biological Engineering, University of Colorado Boulder, Boulder, CO, USA 80309}
\email{michael.shirts@colorado.edu}

\begin{document}

\maketitle

\begin{abstract}
    In reverse osmosis (RO) and nanofiltration (NF) membranes, the polymer structure and interactions with solvent and solutes dictate the permeability and selectivity. However, these interactions have not been fully characterized within hydrated polymer membranes. In this study, we elucidate the local atomic neighborhood around ions within a RO membrane using molecular dynamics (MD). We built a MD model of a RO membrane closely following experimental synthesis and performed long time scale simulations of ions moving within the polymer. We find that the ion-oxygen nearest neighbor distance within the membrane is essentially the same as in solution, indicating that ions coordinate similarly in the confined membrane as in water. However, we do find that the average coordination number decreases in the polymer, which we attribute primarily to shifting the outer portion of the solvation shell beyond the cutoff, rather than being entirely stripped away. We find that cations bind tightly to both the carboxylate and amide oxygen atoms within the membrane. Even in ionized membranes, binding to amide oxygen atoms appears to play a substantial role in hindering ion mobility. Finally, we find that commonly used measures of ionic solvation structure such as coordination numbers do not fully capture the solvation structure, and we explore other measures such as the chemical composition of the nearest neighbors and the radial distribution function.
\end{abstract}

\section{Introduction}

Reverse osmosis (RO) and nanofiltration (NF) membranes are the most widely used technologies for water desalination and purification. These membranes exhibit excellent water-salt selectivity. However, they struggle with more specific separations, such as distinguishing between ions of similar size or charge. These types of separations are increasingly important for applications in resource recovery and energy storage~\cite{foo_solvent-driven_2022, peng_extreme_2024}. The mechanisms governing such complex ion separations are not well understood, and recent research has focused on uncovering the molecular-level processes that drive selective transport~\cite{epsztein_towards_2020, shefer_applying_2022, zuo_selective_2021}. Studying the nanoporous structure of polymer membranes is necessary to fully describe the molecular-level mechanisms that govern transport in these salt-rejecting membranes. The membrane structure and permeant interactions are the primary determinants of transport in polyamide membranes~\cite{epsztein_towards_2020, heiranian_mechanisms_2023, dischinger_unifying_2024}. We must understand these molecular mechanisms to design next-generation water treatment technologies for new applications.

However, typical experimental techniques struggle to characterize the nanoscale structure within polymer membranes, particularly \textit{in operando}~\cite{heiranian_mechanisms_2023, lu_situ_2021}. New research using advanced scanning transmission electron microscopy~\cite{culp_nanoscale_2021, li_three-dimensional_2025}, positron annihilation lifetime spectroscopy (PALS)~\cite{fujioka_probing_2015}, and time-of-flight secondary ion mass spectrometry~\cite{lu_situ_2021} have begun to probe the nanoscale polymer structure and ion coordination in RO membranes, but these techniques struggle to measure the local environment around ions within hydrated membranes. 

Molecular dynamics (MD) simulations provide a route to directly observe these molecular interactions within polymer membranes~\cite{heiranian_molecular_2022, heiranian_mechanisms_2023, freger_solution-diffusion_2024}. While numerous studies have characterized the polymer structure~\cite{ebro_molecular_2013, ridgway_molecular_2017, harder_molecular_2009, ding_structure_2014, muscatello_multiscale_2017, song_molecular_2022, vickers_molecular_2022}, to our knowledge, no MD studies have looked at relevant ions beyond Na$^+$ and Cl$^-$ in simulations of physically relevant RO membranes. Foo et al.~revealed pH-dependent mechanisms of Li$^+$, Mg$^{2+}$, and Cl$^-$ transport through NF membranes~\cite{foo_positivelycoated_2024}, but typical RO membranes are denser and involve different monomer chemistries. For Na$^+$ and Cl$^-$, previous MD studies have identified key factors influencing ion diffusion and partitioning~\cite{kotelyanskii_atomistic_1998, hughes_computational_2010, luo_computer_2011, kolev_molecular_2015, shen_rejection_2016, foo_positivelycoated_2024}. These studies report significantly hindered diffusion coefficients and rejection rates exceeding 99\%, consistent with experimental observations for RO membranes~\cite{kotelyanskii_atomistic_1998, luo_computer_2011, kolev_molecular_2015, he_effect_2023, he_molecular_2023}. Multiple studies reported that the coordination number must decrease for the ions to partition into the membrane and that the ions remained partially dehydrated while in membrane nanopores~\cite{kotelyanskii_atomistic_1998, hughes_computational_2010, shen_dynamics_2016, he_molecular_2023}. By comparing the dynamics of both ions and small organic solutes, Shen et al.~concluded that rejection is largely governed by the dehydration energy barrier rather than the size of the dehydrated solutes~\cite{shen_dynamics_2016}. These studies provide some insight into the mechanisms governing ion rejection in polyamide membranes. However, they do not fully characterize the molecular interactions, and they do not examine many relevant ions. We thus provide a detailed picture of how ions interact with water molecules and the polymer, and we compare a range of relevant cations spanning bare ion sizes and valencies.

To understand the local ion environment within the membrane, we characterize the surroundings of ions using three complementary structural descriptors: radial distribution functions (RDFs), coordination number distributions, and nearest neighbor distributions. Each of these metrics provides distinct insights into the spatial organization and local environment experienced by ions in the membrane.

Radial distribution functions quantify the \textit{average} particle density of relevant chemical groups surrounding an ion, normalized against a homogeneous reference system at large distances. While RDFs effectively illustrate the distribution of possible distances and reveal longer-range structural features beyond the first coordination shell, they do not resolve discrete spatial patterns occurring with individual ions. Instead, all local environments are averaged together, which can obscure the most typical structural motifs when there is multimodal behavior.

Coordination numbers count the number of relevant chemical groups within the ion's first coordination (or solvation) shell. This shell is typically defined by the first minimum in the RDF, which marks the boundary between closely associated groups and the bulk or second shell. However, for some groups, particularly for large ions, the separation between coordination shells is not well-defined. The coordination shells are diffuse and disordered. In such cases, the coordination number becomes sensitive to the determination of this cutoff. Changing the chemical environment from solution to the membrane can change the location of these minima and thus, the integrated number in the shells. The presence of significant density near the RDF minimum can lead to inflated values for the coordination number that include species that are not as relevant to the local interactions as closer species.

Nearest neighbor analysis provides an alternative approach by identifying the heavy atoms closest to the ion, eliminating the dependence on a cutoff radius. This method avoids the arbitrariness of coordination shell boundaries but introduces its own challenges. Specifically, it can be difficult to determine how many of the nearest neighbors are truly relevant to the ion's local environment. For example, considering the eight nearest neighbors of Na$^+$ may include polymer atoms that are nearby due to covalent bonding but would not be included in a physically meaningful, intuitive interaction grouping. In contrast, for Sr$^{2+}$, the eight nearest neighbors typically correspond to a well-defined first solvation shell, consistent with the most common shell in solution. Each of these descriptors has limitations. Therefore, we present the local ion environment by integrating all three metrics.

In this study, in order to gain better insight into the interactions between a membrane and salt solutions, we have constructed a molecular dynamics model of a polyamide RO membrane and characterized the local environment around cations within the membrane. Our membrane model is thick enough to observe bulk, homogeneous membrane properties. It is consistent with experimental measurements of dry and hydrated density, degree of crosslinking, and hydration percentage. We have created neutral and negatively-charged polyamide membranes in order to represent a range of possible operating conditions and to determine the effects of fixed charges on the ion coordination environment. We examined a variety of mobile cations relevant for RO membranes, including Na$^+$, K$^+$, Rb$^+$, Sr$^{2+}$, and Ca$^{2+}$, representing small and large monovalent and divalent cations.

\section{Methods}

\subsection{Polymerization} \label{s:polymerization}

We employed a version of the widely adopted ``progressive crosslinking" approach first introduced by Harder et al.~\cite{harder_molecular_2009, vickers_molecular_2022} to synthesize the highly crosslinked polyamide membrane. This procedure performs crosslinking based on a distance heuristic in an equilibrated ``soup" of randomly packed monomers~\cite{harder_molecular_2009, ridgway_molecular_2017}. We first packed the box and equilibrated with the expected number \textit{m}-phenylenediamine (MPD) monomers and trimesoylchloride (TMC) monomers for a set of physical constraints as described in Supporting Information Section~\nameref{SI:s:monomer_equilibration}. To model a realistic membrane-solution interface, we implemented harmonic walls in the z-dimension of the simulation box. The harmonic walls had a force constant of 10~kcal/mol/\AA$^2$ and a cutoff distance of 10~\AA. When running molecular dynamics with walls, we made the z-dimension non-periodic and performed a slab estimate for long-range electrostatic interactions~\cite{yeh_ewald_1999}. Thus, the walls prevented molecules from spanning the z boundary. 

We crosslinked the monomers to a target degree of crosslinking of 90\%, consistent with experimental values for standard RO membranes~\cite{coronell_depth_2011, karan_sub10_2015}. We performed the polymerization reaction using the REACTER software implemented in LAMMPS~\cite{gissinger_modeling_2017, gissinger_reacter_2020}. The final crosslinked polymer membrane reached 90.9\% crosslinked, 10.0~nm thick, and a density of 0.296~g/cm$^3$. While the membrane achieved the desired degree of crosslinking and thickness, the polymerization was performed for a short time at 300~K. Therefore, the polymer was not able to rearrange and pack to the expected dry density (1.22--1.28 g/cm$^3$)~\cite{karan_sub10_2015}. We achieved correct polymer density in subsequent annealing steps (Section~\nameref{s:polymer_equilibration}). Further details on the polymerization procedure are included in Supporting Information Section~\nameref{SI:s:polymerization}.

We terminated the remaining reactive groups with multiple iterations of inserting free hydroxide (OH) groups and performing termination reactions with REACTER. Once termination was complete, we removed the remaining free OH groups and reassigned partial charges to the polymer. Reassigning the partial charges ensured the system was charge neutral for subsequent equilibration and hydration steps. We provide further details on the termination procedure in Supporting Information Section~\nameref{SI:s:termination}

\subsection{Equilibration} \label{s:polymer_equilibration}

We then performed the annealing procedure from the Polymatic software in order to achieve appropriate polymer densities~\cite{liyana-arachchi_ultrathin_2016, abbott_polymatic_2013}. This equilibration scheme incorporates a series of NVT and NPT steps at high temperatures and gradual compression and decompression. Details for the procedure can be found in Abott et al.~2013~\cite{abbott_polymatic_2013}. We modified the 21 steps in three ways:

\begin{enumerate}
    \item We applied the same harmonic walls that we introduced during polymerization.
    \item All NPT steps were changed to only perform pressure coupling in the x and y dimensions, which ensures we maintain the target membrane thickness (10~nm).
    \item We split the 50~ps, high pressure (30,000~atm) NPT step (step~6) into two steps – 2~ps then 48~ps. Our simulation box changed drastically at these high pressures, causing the domain decomposition grid to change. Splitting this step into two steps allowed us to reassign the processor grid.
\end{enumerate}

\begin{figure}[H]
    \centering
    \includegraphics[width=\textwidth]{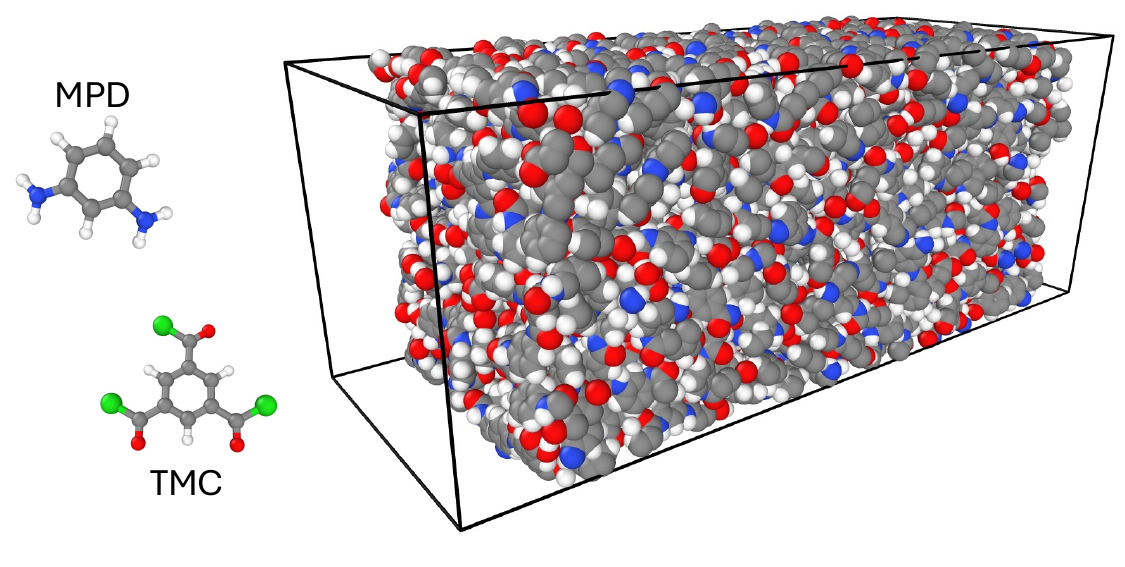}
    \caption{\textbf{Visualization of the membrane monomers and fully equilibrated membrane after polymerization.} The interface (initially created by the harmonic walls, but which remains after polymerization) is clearly shown. The atom colors are as follows: carbon is gray, hydrogen is white, nitrogen is blue, oxygen is red, and chlorine is green.}
    \label{fig:membrane3D_equilibration}
\end{figure}

Finally, we removed any remaining unreacted monomers from the simulation. The final dry density of our simulation was 1.26~g/cm$^3$, which agrees well with previous simulations (1.23-1.29~g/cm$^3$)~\cite{vickers_molecular_2022} and experiment (1.22-1.28 g/cm$^3$)~\cite{karan_sub10_2015}. Figure~\ref{fig:membrane3D_equilibration} depicts the equilibrated membrane.

\subsection{Hydration} \label{s:hydration}

Diffusion into the highly crosslinked polymer membrane is a slow process, so we inserted a fraction of the total water molecules for hydration directly into the membrane to accelerate this membrane hydration process. Most simulation studies of these membranes insert water molecules into the membrane to the desired weight percent water (23~wt\%)~\cite{kotelyanskii_atomistic_1998, shen_dynamics_2016} or force water molecules from a water reservoir into the membrane with a piston~\cite{he_molecular_2023}. However, Vickers et al.~showed that pre-inserting water molecules into the largest voids preserved membrane structure and accelerated the hydration process compared to water molecules diffusing into the membrane~\cite{vickers_molecular_2022}. We iteratively inserted water molecules into the voids in the membrane until the weight percent water was 20~wt\%. Between each iteration, we performed steepest descent minimization to remove any overlapping. To further accelerate the hydration process, we converted from the LAMMPS molecular dynamics engine to GROMACS. Converting to GROMACS improved the simulation speed by over an order of magnitude. We used GROMACS 2023.1 for hydration and ion dynamics simulations~\cite{abraham_gromacs_2015}. 

\begin{figure}[H]
    \centering
    \includegraphics[width=0.8\textwidth]{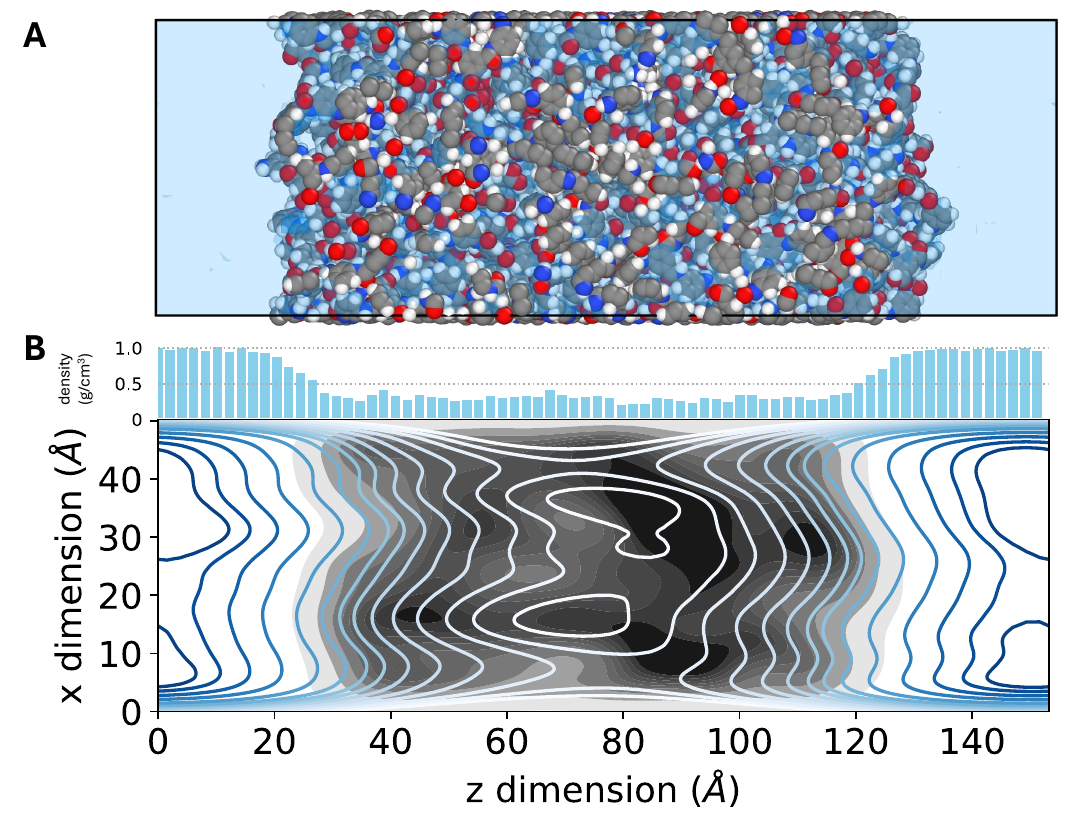}
    \caption{\textbf{Fully hydrated membrane model.} (\textbf{A}) Visualization of the membrane after hydration. The water molecules have been replaced by a light blue surface mesh. The wt\% water is 23\%, and the density of the inner 50\% of the membrane is 1.34~g/cm$^3$. (\textbf{B}) A 2D representation of the membrane with the water density distribution plotted in the marginal axis. The membrane polymer density is plotted as a kernel density estimate (KDE) in the x-z plane in greyscale. The KDE is weighted by the polymer mass, such that the darker regions are the densest parts of the membrane. The water is plotted as contour lines from a KDE on top of the polymer, with darker contour lines representing greater water density. The water mass density profile is shown in the marginal axis. Each bin has a z-width of 2~\AA.}
    \label{fig:membrane_hydration}
\end{figure}

The remaining 3~wt\% was reached by putting the membrane in contact with a water reservoir. We no longer applied the harmonic walls, so the water reservoir was in contact with both membrane interfaces. The resulting hydrated membrane agreed well with reported experimental values. The final wt\% in the inner membrane was 23.03\%~\cite{kotelyanskii_atomistic_1998}. The final membrane dimensions were 5.0 x 5.0 x 10.0~nm$^3$. The final hydrated density was 1.34~g/cm$^3$, within the experimental range 1.30-1.38~g/cm$^3$~\cite{mi_physico-chemical_2007}. The hydrated membrane is visualized in Figure~\ref{fig:membrane_hydration}. In Figure~\ref{fig:membrane_hydration}B, the water density decreases from the bulk value (0.997 g/cm$^3$) in the reservoir to 0.21~g/cm$^3$ at its lowest point in the membrane.

\subsection{Membrane Ionization}

Membrane charge ionization depends on feedwater pH, and during typical operating conditions, at least some of the ionizable moieties introduce fixed charges into the membrane~\cite{ritt_ionization_2020, kimani_influence_2022}. However, the extent that these charges affect ion transport in RO membranes is not well understood. Recent work has reported contradicting arguments--both that membrane charge significantly affects salt rejection~\cite{kimani_influence_2022} and that membrane charge weakly affects salt rejection~\cite{stolov_membrane_2020, chen_facile_2017}. Further, the spatial variation of the fixed charges is not clear. Carboxyl groups show two pK$_a$ values, indicating significant ionization on the surface of the membrane and confined within membrane voids~\cite{coronell_quantification_2008, chen_facile_2017}. However, it has been reported that these interior groups are not ionized at typical operating pH and that spatial heterogeneity creates regions where membrane charge does not affect salt rejection~\cite{ritt_ionization_2020, stolov_membrane_2020}. For this study, we built RO membranes with a range of ionization states. We randomly chose carboxyl groups to ionize, and we ionized 0\%, 25\%, and 50\% of the total carboxyl groups. We chose this range in order to understand how the local ion environment changed within ionized membranes. Ritt et al.~found that for NF membranes, only surface carboxyl groups ionize at neutral pH and interior ionization requires pH $> 9$~\cite{ritt_ionization_2020}. Higher percentages are unlikely to be relevant since the pH would need to be much higher than RO feed streams. Typical desalination processes run around pH 8~\cite{ritt_ionization_2020, pontie_seawater_2013}. We expect the 50\% ionized membrane to be most relevant for standard operating pH, since it is unlikely that all carboxyl groups will be protonated at neutral pH~\cite{ritt_ionization_2020}. We did not protonate any of the amine groups on the polymer. The pK$_a$ of aniline (an amine on a benzene ring) is roughly 4.6, and the electron-withdrawing amide that results from polymerization likely lowers that further.
Therefore, at neutral pH we would not expect the amine groups to be protonated.
Furthermore, the density of amine groups is significantly lower than the carboxyl groups. Further details about the ionization procedure are presented in Supporting Information Section~\nameref{SI:s:ionization}.

\subsection{Ion Insertion and Dynamics}

We inserted cations into the membrane in order to balance the negatively charged polymer and characterize the environment that ions experience in a RO membrane. For the ionized membranes, we inserted enough cations to balance the polymer charge. We inserted cations near the ionized membrane charged groups. Specifically, we replaced the water molecules near the carboxylate group with a cation. For the charge neutral membranes, we inserted the same number of cations necessary for the 50\% ionized membranes. This insertion ensured we could sufficiently sample configurations with ions in the membrane. Rather than ionized carboxylate groups, we placed ions nearby carboxylic acid groups. Adding these cations created a positively charged system, so we balanced the cations by inserting anions into the water reservoir. The final system included the polymer membrane, a water reservoir in contact with the membrane, cations inserted into the membrane, and both cations and anions in the water reservoir. 

To exhaustively sample the local environment ions experience in a membrane, we ran each system for 2000~ns in production NPT. We performed these long time-scale simulations for the salts and ionization states shown in Table~\ref{tab:ion_dynamics_simulations}. All aqueous salt solution simulations were run as described in our previous work~\cite{schwindt_molecular_2025}.

\begin{table}[H]
    \centering
    \caption{\textbf{Salts and ionization states for the long time-scale simulations.}}
    \begin{tabular}{|c|c|}
        \hline
        \textbf{Salt} & \textbf{Ionization States} \\
        \hline
        NaCl & 0\%, 25\%, 50\% \\
        \hline
        KCl & 50\% \\
        \hline
        RbCl & 0\%, 25\%, 50\% \\
        \hline
        CaCl$_2$ & 50\% \\ 
        \hline
        SrCl$_2$ & 0\%, 25\%, 50\% \\
        \hline
    \end{tabular}
    \label{tab:ion_dynamics_simulations}
\end{table}

\section{Results and Discussion} \label{s:results}

We provide a detailed picture of the local environment experienced by cations within our RO membrane simulations using radial distribution functions (RDFs), coordination numbers, and nearest neighbors around the ions. For each of these, we only consider cations that remain in the bulk membrane, defined as the inner 50\% of the polymer.

\subsection{Primary ion-oxygen coordination distances are largely unchanged between solution and the membrane}

\begin{figure}[H]
    \centering
    \includegraphics[width=\textwidth]{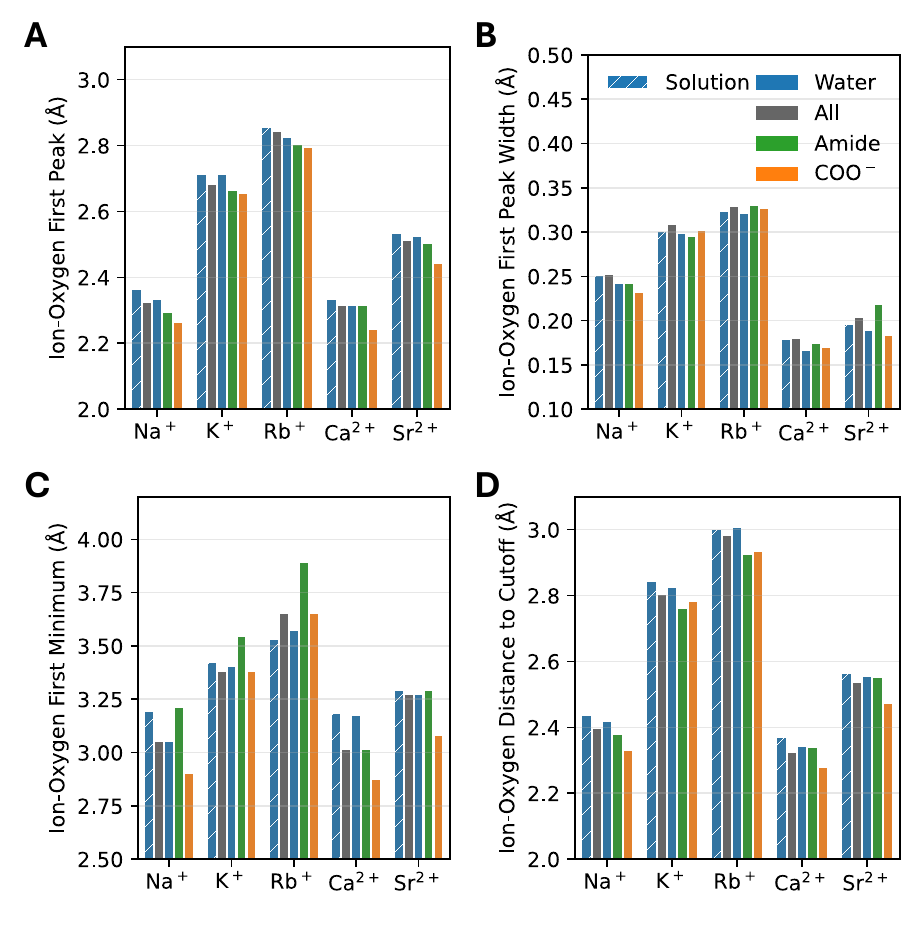}
    \caption{\textbf{Ion-oxygen distances demonstrate how cations bind with different oxygen species in solution and in membrane.} We extracted ion-oxygen distances from the radial distribution functions with bin width 0.01~\AA. We include first peak location (\textbf{A}), first peak width (\textbf{B}), first minimum location (\textbf{C}), and average distance up to the first minimum (\textbf{D}). The hashed bars are for these ions in solution, and thus are always for ion-water oxygen. The gray ``All'' bar corresponds to all oxygen atoms in the system regardless of species or functional group. Peaks must be separated by at least 0.05~\AA~and must be at least 0.1~\AA~in width. The first peak width is the width at half max. The average to cutoff is calculated with all ion-oxygen distances less than the first minimum in the RDF. }
    \label{fig:ion-oxygen_distances}
\end{figure}

Figure~\ref{fig:ion-oxygen_distances} presents four aspects of the ion-oxygen distance distributions for the cations we examined, extracted from the RDFs: (A) first peak location, (B) first peak width, (C) first minimum location, and (D) average ion-oxygen distance up to the first minimum. We define the primary ion-oxygen coordination distances as those that contribute to the first peak in the RDF. The first peak width quantifies the variance in this interaction. The first minimum marks the boundary between the first and second coordination shells. All these quantities follow trends consistent with bare ionic radii, with the ordering Na$^+$ $<$ Ca$^{2+}$ $<$ Sr$^{2+}$ $<$ K$^+$ $<$ Rb$^+$. Across all ions, the first peak distances in solution are slightly larger than those in the membrane, though the magnitude of this difference remains small (Figure~\ref{fig:ion-oxygen_distances}A). The difference in the solution first peak and the membrane first peak for ``All'' oxygens in the system is at most 0.04~\AA. Among oxygen species, carboxylate oxygens consistently exhibit the shortest ion-oxygen distances, implying that ions tightly coordinate with carboxylate groups. The largest difference from solution is for the Na$^+$-carboxylate oxygen distance, which is 0.1~\AA~smaller than the Na$^+$-water distance in solution.

Divalent ions have little variance in the spatial arrangement of coordinating groups, but the larger monovalent ions have more diffuse coordination shells. Ca$^{2+}$ and Sr$^{2+}$ have small first peak widths (Figure~\ref{fig:ion-oxygen_distances}B). In contrast, K$^+$ and Rb$^+$ have larger peak widths, and these widths do not change between solution and membrane. Thus, we observe similar variance in the ion-oxygen distances in the membrane as in solution. However, the high charge density ions (Na$^+$, Ca$^{2+}$, and Sr$^{2+}$) show reduced peak widths in the membrane compared to solution. An exception is observed for Sr$^{2+}$ interacting with amide oxygens, where the peak width is significantly larger than for other Sr$^{2+}$-oxygen interactions. This discrepancy arises because Sr$^{2+}$ does not strongly or frequently interact with amide oxygens. The peak widths for the ion-all oxygen RDFs (regardless of species or functional group) in the membrane are greater than or equal to those in solution, despite each of the individual widths for water, amide, and carboxylate being lower in the membrane. This increased variance is attributed to the heterogeneous chemical environment of the membrane. Specifically, this RDF includes carboxylic acid groups that spread out the first peak.

The high charge density ions coordinate more tightly with carboxylates due to strong electrostatic interactions. The coordination shell cutoff for carboxylate oxygens is smaller than for other oxygen species (Figure~\ref{fig:ion-oxygen_distances}C). The partial charge on the amide oxygen is -0.5851, and the partial charge on an individual carboxylate oxygen is -0.8204. The amide oxygen coordination shell cutoff for monovalent ions is larger, despite similar first peak distances. The first minima are larger because there is no amide oxygen density between the first and second shells in the membrane, and the second shell for the amide oxygen is farther out than those for other oxygen species. For example, Na$^+$ shows an increase in amide oxygen density only beyond 4~\AA~in the membrane, compared to 3.5~\AA~for the water oxygen density as seen in Supporting Information Figures~\ref{SI:fig:membrane_RDFs_2} and~\ref{SI:fig:solution_RDFs}.

Additional oxygen atoms contribute to the coordination shells beyond the primary ion-oxygen distances. When ion-oxygen distances are calculated by averaging all distances up to the first minimum (Figure~\ref{fig:ion-oxygen_distances}D), the values increase across all scenarios compared to the first peak location, particularly for monovalent ions. The weaker electrostatic interactions for monovalent ions result in more diffuse coordination shells with significant density between the first peak and the first minimum. However, all ions demonstrate this behavior to a degree, which highlights that these systems are highly disordered, particularly beyond the primary interactions.

\subsection{Specific ion effects on the speciation of the ion nearest neighbors}

\begin{figure}[H]
    \centering
    \includegraphics[width=\textwidth]{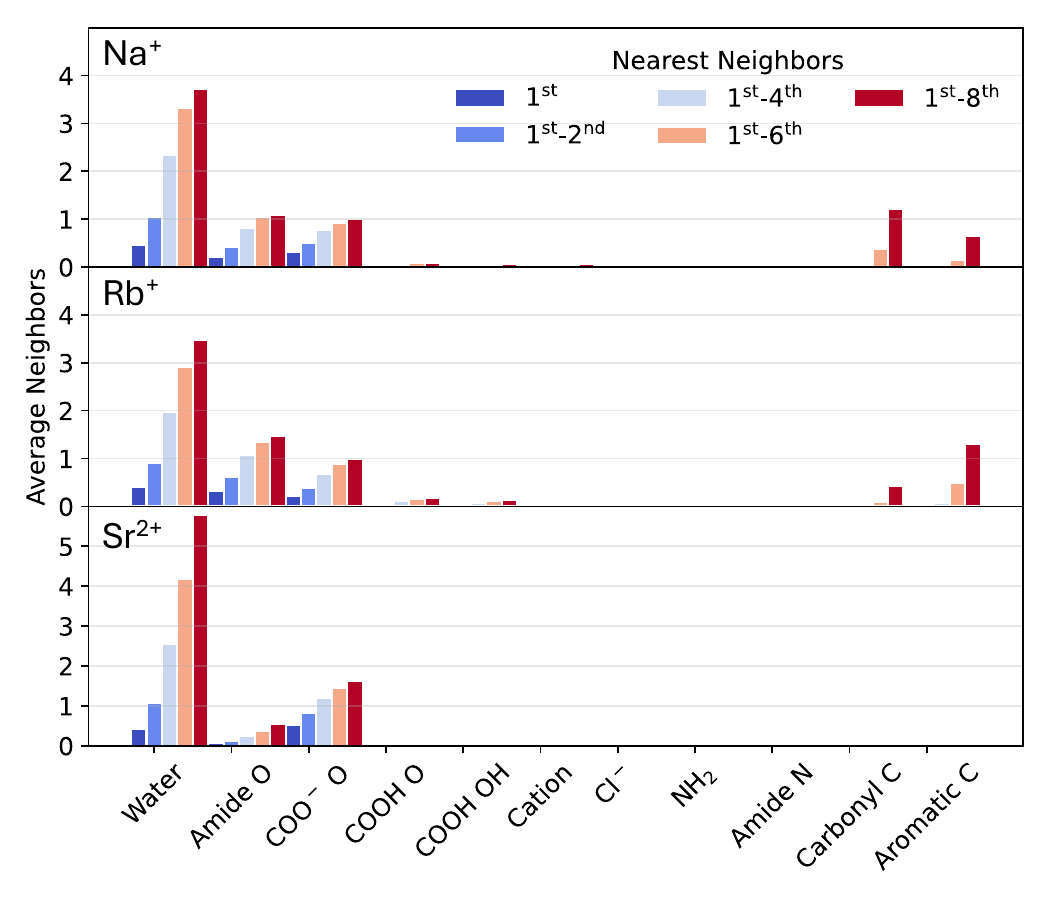}
    \caption{\textbf{The tightly bound nearest neighbors are distributed across water, amide, and carboxylate oxygens, but more distant neighbors can include other polymer groups.} We present distributions of the average number of nearest neighbors for Na$^+$, Rb$^+$, and Sr$^{2+}$. The color indicates the distribution up to the $n$th nearest neighbor, such that the sum across one color will yield $n$. For example, on average for the two nearest neighbors of Na$^+$, there is one water molecule and a 50:50 split of either an amide oxygen or a carboxylate oxygen. For all ions, the first four neighbors are all oxygen atoms, but they are distributed differently among water molecules, amide oxygens, and carboxylate oxygens. }
    \label{fig:neighbor_speciation_Ntot}
\end{figure}

The tightly bound nearest neighbors are always oxygen atoms regardless of ion environment, and they are distributed among water, amide, and carboxylate oxygens depending on ion size and valency. Figure~\ref{fig:neighbor_speciation_Ntot} shows how the nearest neighbors are distributed across various species in the membrane when considering the 1st through 8th nearest neighbors. The first neighbor is one of a water molecule, an amide oxygen, or a carboxylate oxygen. For Na$^+$ and Rb$^+$, the first neighbor is most often a water molecule, while for Sr$^{2+}$, it is typically a carboxylate oxygen, reflecting its stronger electrostatic interactions. Across all three ions, the first two nearest neighbors consistently include one water molecule and either an amide or carboxylate oxygen. For the second neighbor, Rb$^+$ tends to favor amide oxygens over carboxylates, while Na$^+$ and Sr$^{2+}$ favor carboxylate oxygens. The two nearest neighbors are strongly influenced by the charge density of the ions. Rb$^+$ is larger and therefore has a lower charge density. We find that it does not bind as readily with the highly charged carboxylate groups.

Furthermore, these tightly bound nearest neighbors are located at the same distances as water molecules in solution -- out to the 4th or 5th nearest neighbors for Na$^+$ and Rb$^+$ and the 7th or 8th nearest neighbors for Sr$^{2+}$. Figures~\ref{SI:fig:neighbor_distances_Na},~\ref{SI:fig:neighbor_distances_Rb}, and~\ref{SI:fig:neighbor_distances_Sr} in the Supporting Information confirm that the neighbor distance distributions are consistent in both solution and membrane. If we consider only the nearest oxygen atoms, the neighbors beyond the tightly bound atoms are very dispersed and less structured in the membrane. For example, for Na$^+$, the 6th through 8th nearest neighbors appear anywhere between 2.5 and 5~\AA~from the ion. Figure~\ref{SI:fig:neighbor_distances_Na} reveals a small amount of structure in the 7th and 8th neighbors, but the distributions are much broader than in solution. These neighbors correspond to the second solvation shell, which becomes increasingly disordered in the membrane due to spatial constraints. The polymer network disrupts the second solvation shell shifting this density farther from the ion. The polymer forms voids 4-5~\AA~in diameter, which is similar to where the second solvation shells forms in solution.

More than one polymer atom enters the ion neighborhood only beyond the 4th nearest neighbor. Na$^+$'s nearest neighbors never include more than one amide oxygen or carboxylate oxygen. Even when considering up to the 8th nearest neighbor, the average number of amide oxygens and carboxylate oxygens remains at 1, as seen with the red bars in the top panel of Figure~\ref{fig:neighbor_speciation_Ntot}. Instead, the nearest neighbors include carbon atoms from the polymer that are covalently bonded to the oxygen atoms. However, the larger Rb$^+$ and Sr$^{2+}$ frequently are near two amide or carboxylate oxygens, respectively. These ions can interact with two different monomer fragments on the polymer, but these interactions are longer range -- only occurring in the 5th to 8th nearest neighbors. For Sr$^{2+}$, the polymer interactions could be within 3~\AA~of the ion, but for Rb$^+$, they could be out to 5~\AA. Notably, Sr$^{2+}$ only ever includes oxygen atoms in its 8 nearest neighbors, which is consistent with its behavior in solution. In solution, Sr$^{2+}$ is tightly coordinated with 8 water molecules, but Na$^+$ and Rb$^+$ have coordination numbers below 8.

\subsection{Ions in the membrane remove multiple water molecules from their solvation shells and replace them with 1-2 polar polymer groups}

\begin{figure}[H]
    \centering
    \includegraphics[width=\textwidth]{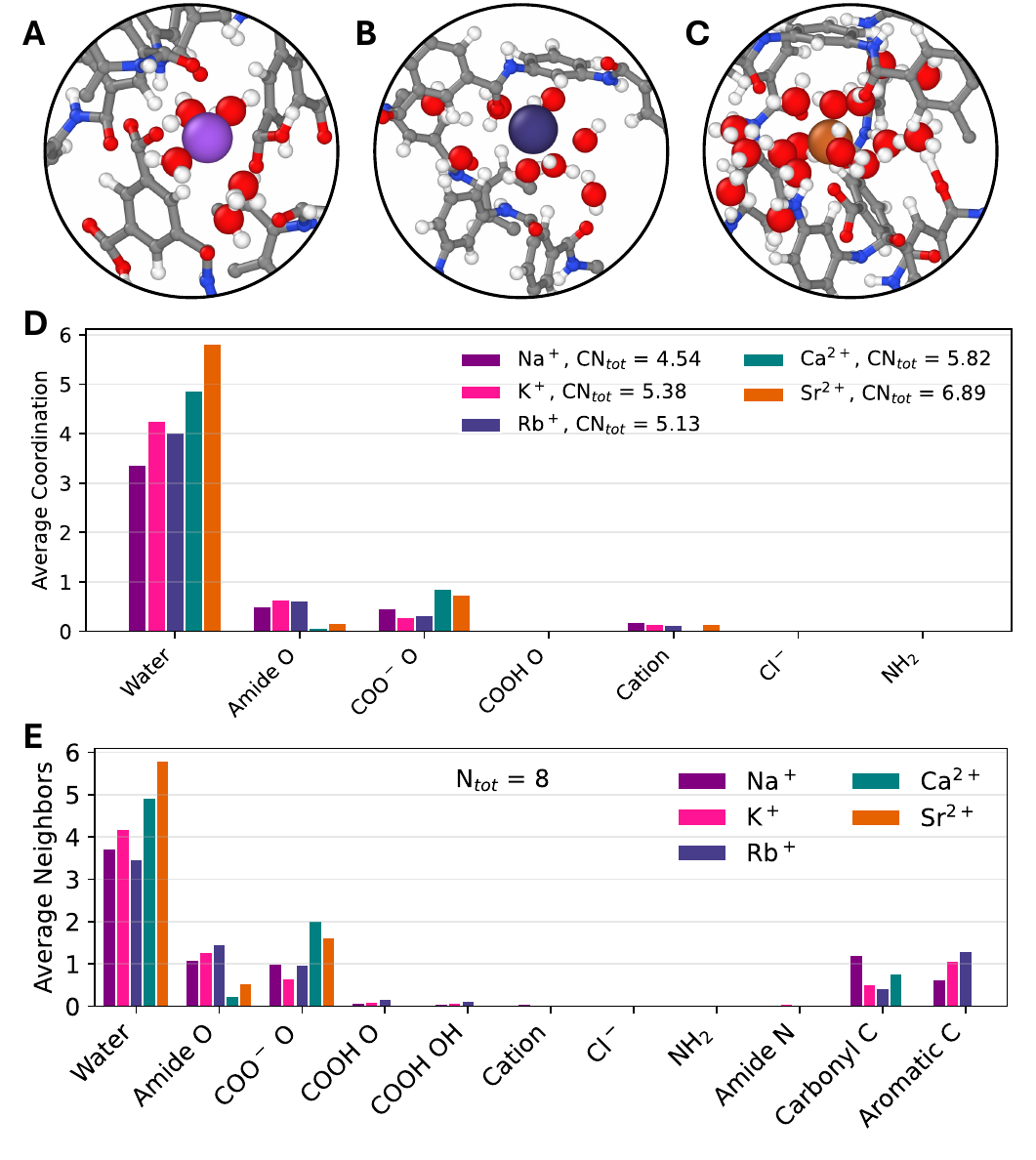}
    \caption{\textbf{Ions bind tightly to carboxylate groups and amide oxygens in the membrane, and they maintain partial hydration shells.} We show representative snapshots of Na$^+$ (A), Rb$^+$ (B) and Sr$^{2+}$ (C) in the 50\% ionized membrane and distributions of the average coordination shells (D) and the 8 nearest neighbors (E) in the membrane for all ions. Coloring in the visualization follows Figure~\ref{fig:membrane3D_equilibration}. We include the carboxylic acid group, other cations, anions, and the amine group to emphasize that we see little to no interactions between these species and the cations. The total coordination number for each ion is given in the legend of (D). These total coordination numbers are distributed according to the plot.}
    \label{fig:ion_speciation_and_shells}
\end{figure}

Compared to solution, the total coordination number for an ion is consistently 1–2 units lower in the membrane. Specifically, the coordination number in the membrane decreases by 1.11 for Na$^+$, 1.47 for K$^+$, 2.26 for Rb$^+$, 1.71 for Ca$^{2+}$, and 1.11 for Sr$^{2+}$ from the solution. The average coordination numbers in solution are presented in Supporting Information Figure~\ref{SI:fig:average_coordination_solution}. As seen in the legend for Figure~\ref{fig:ion_speciation_and_shells}D, the total coordination number for Rb$^+$ is smaller than for K$^+$. This trend is reversed in solution, where Rb$^+$ is larger by about 0.5. Rb$^+$ coordination shell is larger than the other ions, and thus the confinement imposed by the polymer forces Rb$^+$ to a lower coordination number.

Despite the reduction in coordination number, ions remain predominantly hydrated in the membrane, with some coordination from polymer functional groups. Figure~\ref{fig:ion_speciation_and_shells}D details how the coordination shell in the membrane is distributed across atomic species for all ions studied. As discussed for nearest neighbors above, we find that the lower coordination number pushes some density outside the coordination shell, rather than fully stripping the species away. Additionally, the ions bind closely with the polymer, which introduces a steric penalty to fill the remainder of the coordination shell.

We only observe significant coordination with water molecules, amide oxygens, and carboxylate oxygens. The small amounts of cation-cation coordination are due to doubly coordinated carboxylate groups. Supporting Information Figure~\ref{SI:fig:cation-cation_coordination} shows example snapshots of this behavior for Na$^+$ and Rb$^+$. Since there are no short range interactions between the cations, the first minimum in the cation-cation RDF is larger than a typical coordination shell. For example, the Na$^+$-Na$^+$ coordination shell cutoff is 4.6~\AA, much larger than the ion-oxygen cutoffs shown in Figure~\ref{fig:ion-oxygen_distances}C.

Surprisingly, monovalent ions more frequently coordinate with amide oxygens than carboxylate oxygens. This result is unexpected since the carboxylate oxygens have higher partial charges than amide oxygens, but it highlights the importance of polar groups, regardless of ionization. This preference for amide coordination is not solely due to the larger coordination cutoff for amide oxygens (Figure~\ref{fig:ion-oxygen_distances}C), as it is also observed in the nearest neighbor analysis (Figure~\ref{fig:ion_speciation_and_shells}E). Additionally, the high amide-to-carboxylate ratio implies in the polymer that there is an entropic bias towards the amide oxygens. The ratio of amide oxygens to carboxylate oxygens in the 50\% ionized polymer is 10.3.


Representative snapshots of Na$^+$ (A), Rb$^+$ (B), and Sr$^{2+}$ (C) illustrate how the larger polymer structure corresponds to the local environment. For example, Figure~\ref{fig:ion_speciation_and_shells}A shows a Na$^+$ ion coordinated by three water molecules and one carboxylate oxygen, totaling four coordinated species. In solution, Na$^+$ typically coordinates with six water molecules, so the ion loses three waters and substitutes one with a polymer oxygen. While not within the coordination shell, Na$^+$ in this snapshot is near a carboxylic acid group and an amide oxygen. The ion's mobility is likely hindered by a combination of these groups--switching between coordination with the carboxylate and amide oxygens while avoiding the bulkier carboxylic acid. Figure~\ref{fig:ion_speciation_and_shells}B shows Rb$^+$ coordinated by four water molecules and one amide oxygen. No carboxylate groups are present in this snapshot, despite initial insertion near ionized groups. In contrast, Sr$^{2+}$ (Figure~\ref{fig:ion_speciation_and_shells}C) draws in many water molecules. This snapshot includes six coordinated water molecules and one carboxylate oxygen, preserving much of its first and even second hydration shell.

Divalent ions also retain more water molecules in their coordination shells as shown in Figure~\ref{fig:ion_speciation_and_shells}D, which can lead to membrane swelling. For instance, the presence of Ca$^{2+}$ increases membrane volume by 4.0\% compared to Na$^+$, despite being similarly sized ions. This artifact would be most relevant for membranes that are slightly negatively charged. The fixed charges would facilitate trapped divalent cations, without completely rejecting the higher valence ions by strong dielectric and Donnan exclusion at the interface~\cite{freger_dielectric_2023}. Additionally, the smaller divalent ion Ca$^{2+}$ includes two carboxylate oxygens and the carbonyl carbon atom among its eight nearest neighbors, while Sr$^{2+}$ never includes carbon atoms and only occasionally coordinates with two carboxylate oxygens (Figure~\ref{fig:ion_speciation_and_shells}E). Sr$^{2+}$ holds more water molecules than Ca$^{2+}$, but its bare ion size is larger. The membrane volume swells less with trapped Sr$^{2+}$ than with Ca$^{2+}$. The volume swells by 2.5\% with Sr$^{2+}$ compared to Na$^+$.

\subsection{The most common coordination shells in the membrane involve one polymer group and multiple water molecules}

\begin{figure}[H]
    \centering
    \includegraphics[width=\textwidth]{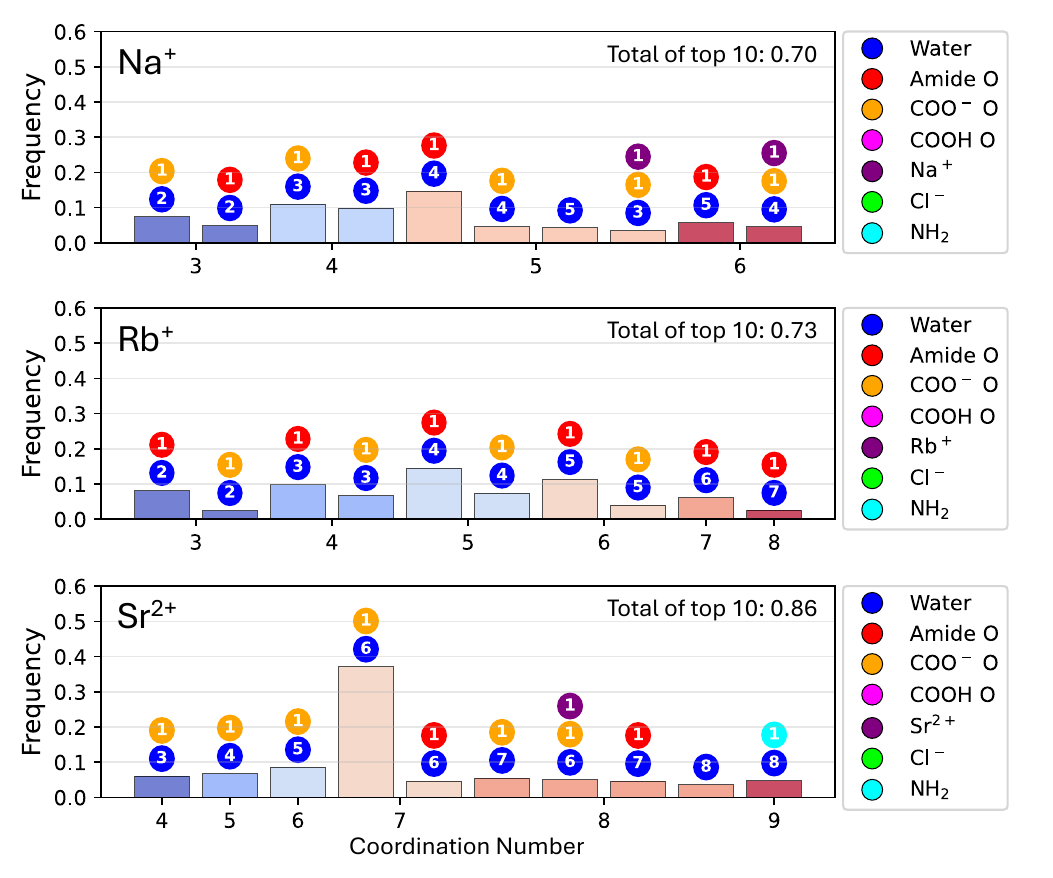}
    \caption{\textbf{The most common shells are 1 amide oxygen and 4 water molecules for both monovalent ions and 1 carboxylate oxygen and 6 water molecules for Sr$^{2+}$, and ions do not coordinate with more than one polymer group at once.} We show distributions across species for the 10 most frequent coordination shells for Na$^+$, Rb$^+$, and Sr$^{2+}$ in the 50\% ionized membrane.  Circles above each bar speciate the shells with the number of coordinating species shown inside the circle. We include all species that we considered in the legend to highlight the lack of some species, specifically anions and carboxylic acid oxygens. The total fraction of the simulation represented by the 10 most frequent shells is included in the top right corner of each panel. The cutoffs for each coordinating species were determined from the first minimum in the membrane cation-species RDFs (Supporting Information Section~\nameref{SI:s:membrane_RDFs}). Bars are colored by the total coordination number. }
    \label{fig:coordination_frequency}
\end{figure}

Ions do not coordinate with more than one polymer group at once. Figure~\ref{fig:coordination_frequency} demonstrates how the ten most common coordination shells are distributed across species for three representative ions—Na$^+$, Rb$^+$, and Sr$^{2+}$. The most frequent coordination shells in the membrane consistently involve one polymer oxygen and multiple water molecules. None of these shells include two polymer groups, which would be indicated by a 2 in the orange or red circles or both an orange and red circle in the same motif. The only shells with more than two distinct species include a coordinated cation, which is primarily an artifact of the cutoff determination as discussed below. For example, the average coordination shell for Na$^+$ (Figure~\ref{fig:ion_speciation_and_shells}D) suggests equal likelihood of coordinating with amide and carboxylate oxygens, but this average shell is not actually sampled. Rather, shells with either one amide oxygen or one carboxylate oxygen appear with similar frequency, indicating that averaging obscures the true coordination shells the ion experiences.

A key limitation of this analysis is its sensitivity to the chosen cutoff distance. These shells are determined by the first minima in the RDFs for ions in the membrane and the respective species. If relevant neighbors lie just beyond the RDF-defined cutoff, they are excluded from the coordination shell, resulting in artificially low coordination numbers. Conversely, species that are not tightly bound but fall just within the cutoff may inflate the coordination number. Furthermore, using the first minimum in the RDF can result in coordination shells that are too large. Species that are not tightly coordinated but do have longer range structure would show a first minimum at a distance beyond a typical coordination shell. For example, the Sr$^{2+}$–amide cutoff used in Figure~\ref{fig:coordination_frequency} is 7.53~\AA, which is significantly larger than typical coordination distances. These extended cutoffs include unexpected species, many of which do not appear among the eight nearest neighbors (Figure~\ref{fig:ion_speciation_and_shells}E).

For both Na$^+$ and Rb$^+$, the most common shell contains four water molecules and one amide oxygen. However, the distributions are balanced, such that many coordination shells have similar frequencies. The local environment for these monovalent ions appears to fluctuate substantially. These results suggest that Na$^+$ and Rb$^+$ jump between nearby amide and carboxylate oxygens while also shedding and gaining water molecules. In contrast, Sr$^{2+}$ shows a dominant coordination shell consisting of six water molecules and one carboxylate oxygen, which occurs in nearly 40\% of the simulation frames. This configuration yields a coordination number of seven, slightly lower than the most frequent shell in solution, which typically includes eight coordinated species. While there may be exchange of water molecules, the local environment around Sr$^{2+}$ does not appear to vary much in the membrane.

\subsection{Effects of polymer ionization and choice of water molecule} \label{s:ionization_and_water_models}

\begin{figure}[H]
    \centering
    \includegraphics[width=\textwidth]{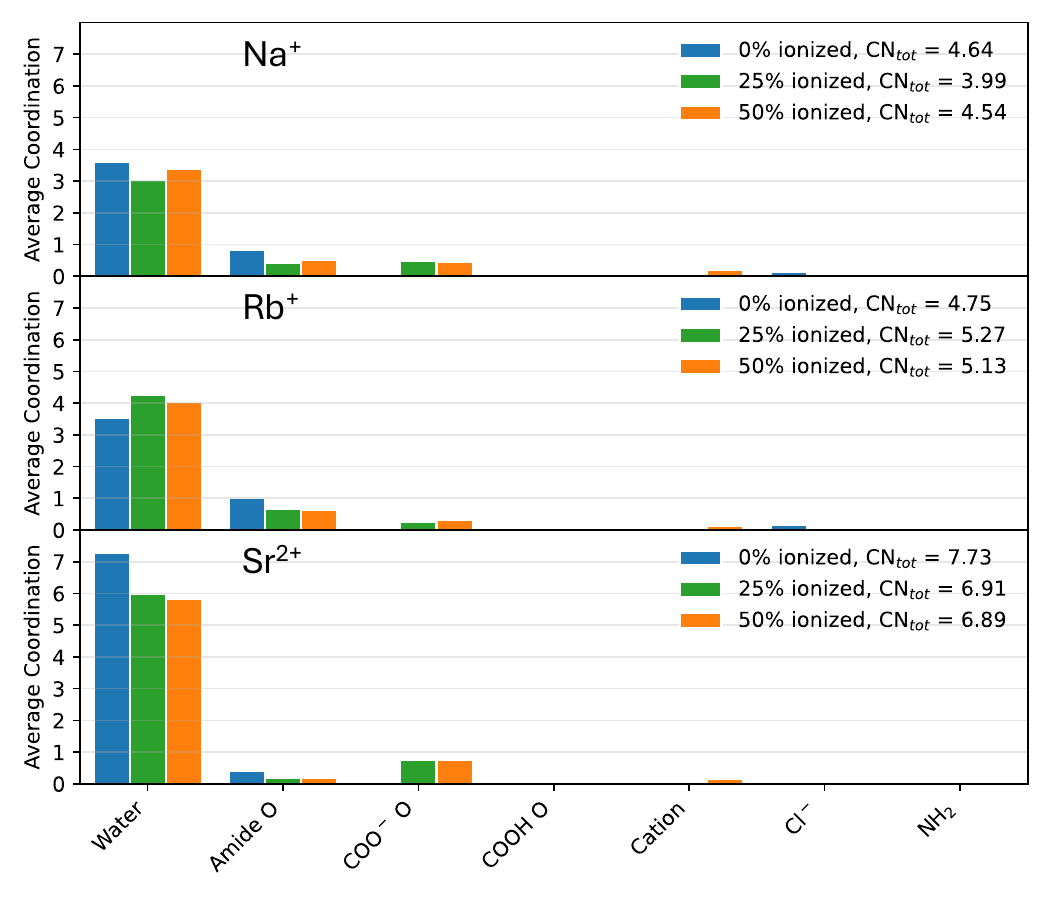}
    \caption{\textbf{The local coordination environment does not differ significantly between ionized membranes. } We plot the distributions of the average coordination shells for Na$^+$, Rb$^+$, and Sr$^{2+}$ for the three membrane ionization states considered. The total coordination number is included in the legend for each panel. }
    \label{fig:coordination_ionization}
\end{figure}

We observed distinct environments for the 0\% ionized, charge-neutral membrane, but we found minimal differences in the local environment across the two ionized membranes. Figure~\ref{fig:coordination_ionization} depicts the average coordination number distributions for Na$^+$, Rb$^+$, and Sr$^{2+}$ across the three ionization levels. Notably, we only observed significant interactions with oxygen atoms in all simulations. For the charge-neutral membranes, the ions coordinate more with the amide oxygens, since they have the most negative partial charge on the polymer. The oxygen atoms in the carboxylic acid groups are similarly charged to the amide oxygen but are blocked by the positively charged hydrogen, geometrically impeding interactions with ions. Na$^+$ and Sr$^{2+}$ with high charge density additionally coordinate with more water molecules in the charge-neutral membrane. We do not observe a clear trend in the total coordination number upon ionization. For Sr$^{2+}$, the total coordination decreases as the polymer is more ionized. The carboxylate fixed charges appear to displace water molecules from the coordination shell compared to the charge neutral membrane.

Rb$^+$ shows little coordination with carboxylate oxygens regardless of ionization. As a result, Rb$^+$ coordination is dominated by water and amide oxygens. Even in ionized membranes, interactions with the amide oxygen play a substantial role in hindering ion mobility by tightly coordinating with cations. Thus, modifying the crosslinking chemistry could influence membrane performance by introducing novel molecular interactions, in addition to changing the membrane density and degree of crosslinking.

We investigated to what extent these results may be dependent on the force field by simulating with two common three point water models and their optimized ion parameters. We examined TIP3P and OPC3 (Li \& Merz 2015~\cite{li_systematic_2015} ions for TIP3P and Sengupta et al.~2021~\cite{sengupta_parameterization_2021} ions for OPC3). Both models capture the essential structural features of ion coordination, but TIP3P exhibits slightly less ion-water interactions than OPC3. Consequently, ions simulated with TIP3P more frequently coordinate with polymer groups. We also observed more cation-cation coordination with OPC3 than with TIP3P, which would arise primarily from differences in the ion parameters rather than the water models.

 \begin{figure}[H]
    \centering
    \includegraphics[width=\textwidth]{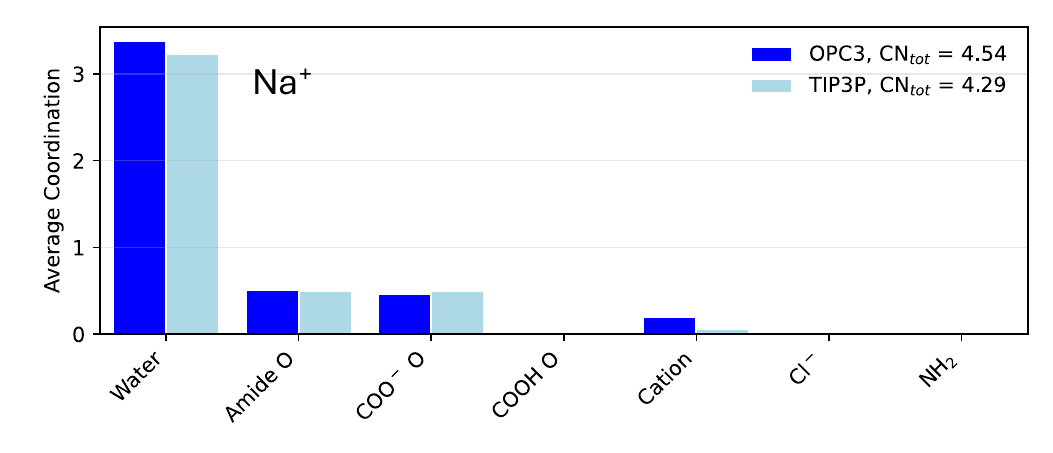}
    \caption{\textbf{We observe minimal differences in the coordination environments for OPC3 and TIP3P water.} We include the distributions of the average coordination shells for Na$^+$ for the two water models considered. The total coordination number is included in the legend. }
    \label{fig:coordination_water_models}
\end{figure}

\section{Conclusions}

In this study, we have built a model of an RO membrane that is maximally consistent with experimental membranes, and we provided a detailed characterization of the local ion environment within polymer membranes. We analyzed radial distribution functions, coordination number distributions, and nearest neighbor distributions to provide a complete picture of the chemical environment that ions experience inside the membrane.

Ions coordinate with oxygen atoms in the membrane in nearly the same way as in solution. The primary ion-oxygen coordination distances are very similar in solution and membrane environments. However, the second solvation shell is diffuse and disordered in the membrane. Ions bind tightly to polymer groups, such that the polymer obstructs the second shell. The membrane voids are similar size as the second solvation shell, which prevents a strongly ordered second shell.

The number of coordinated water molecules decreases by 2-4 in the membrane compared to solution, but the total coordination number only decreases by 1–2 units in the membrane. The ion partially compensates for the water molecules by binding with polar atoms on the polymer, specifically carboxylate and amide oxygens. The decrease in total coordination number typically involves pushing density outside the cutoff, rather than fully removing the local density around the ion. We observed substantial coordination with amide oxides, in particular for monovalent ions. While these monovalent ions balance the fixed negative charges on the polymer, they do not need to remain strongly bound to the carboxylate groups. Rather they show substantial mobility, interacting with other parts of the polymer. Divalent ions bind more tightly and consistently to carboxylate groups, but the higher charge would also increase rejection at the membrane interface.

The local environment around monovalent ions varies significantly, such that many different types of coordination shells are sampled frequently. This heterogeneity implies that monovalent ions jump between polymer groups. However, divalent ions like Sr$^{2+}$ sample fewer coordination shells. The local environments that divalent ions experience does not vary much in the membrane.

\section*{Author Contributions}
\noindent \textbf{Nathanael S. Schwindt:} Conceptualization, Formal Analysis, Methodology, Software, Investigation, Visualization, Writing - Original Draft, Writing - Review \& Editing \textbf{Anthony P. Straub:} Writing - Review \& Editing, Funding acquisition \textbf{Michael F. Toney:} Writing - Review \& Editing \textbf{Michael R. Shirts:} Conceptualization, Resources, Writing - Review \& Editing, Supervision, Funding acquisition

\section*{Supporting Information}
Details and figures on the simulation procedure, including parameterization, monomer equilibration, polymerization, termination, hydration, and ionization. Figures and discussion validating methodology with experimental results. Full tables of the ion-oxygen distances extracted from RDFs. Figures depicting the distribution of nearest neighbor distances and coordination number distributions in solution. Figures for all RDFs from which distances and coordination numbers were extracted, both in solution and in the membrane. Snapshots from the simulation illustrating cation-cation coordination (PDF).

\section*{Acknowledgements}
This material is based upon work supported by the National Science Foundation under Grant No. CBET-2136835 and the United States-Israel Binational Science Foundation (BSF), Jerusalem, Israel (grant No. 2021615). Acknowledgment is made to a GAANN (Graduate Assistance in Areas of National Need) Department of Education fellowship. This work used Bridges-2 at Pittsburgh Supercomputing Center through allocation chm230020p from the Advanced Cyberinfrastructure Coordination Ecosystem: Services \& Support (ACCESS) program, which is supported by National Science Foundation grants \texttt{\#}2138259, \texttt{\#}2138286, \texttt{\#}2138307, \texttt{\#}2137603, and \texttt{\#}2138296. We thank Sasha Neefe (CU Boulder Chemical \& Biological Engineering) for useful discussions on characterizing the local environment.


\pagebreak

\setcounter{figure}{0}
\setcounter{equation}{0}
\setcounter{page}{1}
\setcounter{section}{0}
\captionsetup[table]{labelfont={bf}, labelsep=period, name={Table}}

\renewcommand\thesection{S\arabic{section}}
\renewcommand\thefigure{S\arabic{figure}}
\renewcommand\thetable{S\arabic{table}}
\renewcommand\theequation{S\arabic{equation}}
\renewcommand{\thepage}{S\arabic{page}}

\section*{\Large Supporting Information for ``Local ion environment in polyamide membranes revealed by molecular dynamics''}

\section{Additional simulation details}

\subsection{Parameterization}

We built the polyamide monomers using Avogadro version 1.2 and used \texttt{moltemplate} to assign GAFF parameters and to prepare input parameters for LAMMPS~\cite{hanwell_avogadro_2012, jewett_moltemplate_2021, wang_development_2004, thompson_lammps_2022}. We built the polyamide membrane using \textit{m}-phenylenediamine (MPD) and trimesoyl chloride (TMC), common monomers for RO membranes. The monomers are visualized in Figure~\ref{fig:membrane3D_equilibration}. We cleaned the input data and parameter files with AutoMapper to only include the necessary GAFF parameters~\cite{bone_automapper_2022}. We tested two water models: TIP3P~\cite{jorgensen_comparison_1983} and OPC3~\cite{izadi_accuracy_2016}. TIP3P is a fast and common water model, albeit with known issues. OPC3 is another three-point water model that has been optimized to better reproduce the properties of bulk water. Notably, OPC3 better predicts the radial distribution function (RDF) of bulk water than TIP3P~\cite{kadaoluwa_pathirannahalage_systematic_2021}. We used ion parameters optimized for the different water models. Li \& Merz 2015 ion parameters were optimized for TIP3P. Sengupta et al.~2021 ion parameters were optimized for OPC3~\cite{sengupta_parameterization_2021}. We performed membrane hydration with TIP3P, then converted to OPC3. We compared the local ion environment with TIP3P and OPC3 for Na$^+$ as discussed in Section~\nameref{s:ionization_and_water_models}. Based on this comparison, we performed simulations with all other ions with OPC3.

\subsection{Monomer equilibration} \label{SI:s:monomer_equilibration}

To pack the box, we determined the number of expected water molecules, MPD monomers, and TMC monomers by solving a system of equations from a set of physical constraints. For our first constraint, we targeted a membrane with dimensions 5x5x10 nm$^3$. 10~nm is near the lower limit for industry RO active layers synthesized with the standard interfacial polymerization procedure~\cite{he_molecular_2023, karan_sub10_2015, jiang_water_2018}. Then, we used a 3:2 ratio of MPD and TMC monomers, which is stoichiometric in the number of reactive sites and follows experimental estimates and previous simulation values~\cite{wang_water_2023, vickers_molecular_2022, zhang_molecular_2019}. For our third constraint, we targeted 23~wt\% water, a typical hydration level target for MD simulations of RO membranes based on a commercial reference reported in Kotelyanskii et al.~\cite{vickers_molecular_2022, zhang_molecular_2019, kotelyanskii_atomistic_1998}. Finally, the target hydrated polyamide density was set to 1.38~g/cm$^3$, consistent with the experimental range for RO membranes (1.30-1.38~g/cm$^3$)~\cite{mi_physico-chemical_2007}. We chose the upper bound of the experimental density to ensure we had enough monomers to achieve our desired membrane thickness. We would rather build a membrane that is thicker than the target 10~nm than thinner. These constraints resulted in 734 MPD monomers and 490 TMC monomers, which we packed randomly into a box with dimensions 5x5x11 nm$^3$ using Packmol~\cite{martinez_packmol_2009}.

To equilibrate the monomer system, we ran energy minimization and 5~ns of equilibration on the box of monomers. These simulations were run with LAMMPS (Release LAMMPS/14Dec21-intel)~\cite{thompson_lammps_2022}. We alternated 0.5~ns canonical ensemble (NVT) and isobaric-isothermal ensemble (NPT) simulations to ensure the monomers were at equilibrium before polymerization. We ran steepest descent minimization and used a 0.002~ps timestep for both NVT and NPT steps. The temperature rescale thermostat was used for NVT at 300~K, and the Nos\'e-Hoover thermostat and barostat (damping 0.1~ps for both temperature and pressure) were used for NPT at 300~K and 1~atm. Pressure coupling was only applied in the z-dimension, such that the thickness of the membrane would adjust to reach atmospheric pressure. After equilibration, the thickness of the membrane monomers was 10~nm. To model a realistic membrane-solution interface, we implemented harmonic walls in the z-dimension of the simulation box. The harmonic walls had a force constant of 10~kcal/mol/\AA$^2$ and a cutoff distance of 10~\AA. When running molecular dynamics with walls, we made the z-dimension non-periodic and performed a slab estimate for long-range electrostatic interactions~\cite{yeh_ewald_1999}. Thus, the walls prevented molecules from spanning the z boundary. However, after equilibration, some molecules crossed the z boundary. Therefore, we unwrapped those molecules, which increased the membrane thickness to 11.2~nm.

\subsection{Polymerization} \label{SI:s:polymerization}

To accelerate the polymerization process and to enable reaching the high crosslinking percentage necessary for RO membranes, we compressed the membrane monomers before polymerization. We incrementally compressed the monomers by moving the harmonic walls inward until the slab of monomers was 10~nm thick again. After each compression, we performed steepest descent energy minimization and a short (50~ps) NVT step. For these NVT runs, we used a 0.001~ps timestep and the canonical sampling velocity rescale (v-rescale) thermostat with a damping constant of 0.1~ps~\cite{bussi_canonical_2007}. These simulations removed any excess kinetic energy that resulted from pushing the monomers closer. The resulting system had a dry density of 1.40~g/cm$^3$, slightly higher than the target hydrated density range from experiment (1.30-1.38 g/cm$^3$)~\cite{mi_physico-chemical_2007}. These discrepancies were corrected in subsequent polymerization (Section~\nameref{s:polymerization}), annealing (Section~\nameref{s:polymer_equilibration}), and hydration (Section~\nameref{s:hydration}) steps. Finally, we re-equilibrated the system after compression with walls. We ran energy minimization, a 25~ps NVT simulation, and a 25~ps NPT simulation. NVT was again run with a 0.001~ps timestep and the v-rescale thermostat with a damping parameter of 0.1~ps at 300~K. NPT was also run with a 0.001~ps timestep, but we used the Nos\'e-Hoover thermostat and barostat with both damping parameters set to 0.1~ps. Pressure coupling for the x and y dimensions was coupled, but no pressure coupling was applied to the z dimension. 

Pre- and post-reaction templates for the REACTER software were generated with the AutoMapper tool~\cite{bone_automapper_2022}. These templates defined the geometry and atom types of reactive sites and products of the reaction. Bonds were formed every 0.5~ps by a distance heuristic with a 5~\AA~cutoff. An acyl carbon and an amine nitrogen formed an amide bond, and chlorine and one amine hydrogen were removed. Polymerization steps were conducted in the NVT ensemble with a timestep of 0.001~ps and v-rescale temperature coupling (damping = 0.1~ps). We ran the polymerization step for 10~ns, and the degree of crosslinking at the end was 87.7\%, nearly at the 90\% target.

To further facilitate crosslinking, we relaxed the membrane by incrementally moving the walls outward. Each increment was 2.5~\AA~for each wall, so the z-dimension increased by 5~\AA~on each step. Then we performed energy minimization and 50~ps NVT in the same way as in the compression increments. This procedure was repeated until the membrane was 10~nm thick. We re-equilibrated the system after decompression with the same energy minimization, NVT, and NPT runs as after compression. The resulting density was 0.300~g/cm$^3$, which is similar to the initial packing density of previous simulations~\cite{liyana-arachchi_ultrathin_2016}. To achieve the target degree of crosslinking (90\%), we ran a second polymerization step for another 10~ns. The final crosslinked polymer membrane reached 90.9\% crosslinked, 10.0~nm thick, and a density of 0.296~g/cm$^3$.

\subsection{Termination} \label{SI:s:termination}

We terminated the remaining reactive groups with multiple iterations of inserting free hydroxide (OH) groups and performing termination reactions with REACTER. We created an OH reaction template with GAFF parameters using AutoMapper and \texttt{moltemplate}~\cite{bone_automapper_2022, jewett_moltemplate_2021}. We then placed five OH groups near each remaining chlorine on the TMC monomers and fragments using the LAMMPS \texttt{create\_atoms} command. We ran minimization and a 100~ps termination step. During the termination step, we formed bonds every 0.01~ps with a heuristic distance cutoff of 5~\AA. We repeated the insertion, minimization, and termination procedure until all reactive groups were terminated. Once termination was complete, we removed the remaining free OH groups and reassigned partial charges to the polymer. 

Charge assignment was based on monomer fragments with partial charges assigned by AM1BCC ELF10 from OpenEye. The monomer fragments corresponded to how many terminated groups remained. Therefore, MPD had two possible fragments: terminated or crosslinked. TMC had three possible fragments: terminated, linear, or crosslinked. We modified the partial charges slightly for some fragments in order to maintain charge neutrality for the whole polymer. We determined what the total charge of each fragment should be by applying the following constraints:

\begin{itemize}
    \item Two terminated MPD fragments had to balance a single linear TMC fragment.
    \item A single crosslinked MPD fragment had to balance two terminated TMC fragments.
    \item Three terminated MPD fragments had to balance one crosslinked TMC fragment.
    \item Finally, the total charge must be neutral.
\end{itemize}

Under these constraints, the most we needed to modify the AM1BCC charges was 0.01326 on crosslinked TMC fragments. We applied the small charge corrections to aromatic carbons to minimize physical differences. Notably, no modifications were necessary for crosslinked MPD fragments (the most common fragment) or terminated TMC.

\subsection{Hydration} \label{SI:s:hydration}

The final hydration step was performed by placing the membrane in contact with a water reservoir. The water reservoir contained 5046 water molecules, which corresponded to $\frac{2}{3}$ of the membrane thickness. The diffusive hydration process involved an iterative procedure of NPT steps. First, we performed energy minimization and 500~ps of NVT equilibration on the partially hydrated membrane and water box. We used a 0.002~ps timestep and the v-rescale thermostat (\texttt{tau\_t} = 0.1~ps) at 300~K. We then ran 10~ns NPT steps at 1~atm until the 23~wt\% water was achieved. The NPT steps were run with a 0.002~ps timestep, the velocity rescale thermostat (\texttt{tau\_t} = 0.1~ps), and exponential relaxation pressure coupling, or C-rescale (\texttt{tau\_p} = 10~ps). For this final hydration step, we calculated the weight percent water for the “bulk” membrane, defined as the innermost 50\% of the polymer. This procedure required 120~ns to reach the desired hydration percentage. The hydration process was carried out with TIP3P water, but we tested the robustness of our simulations by comparing with OPC3 water. To hydrate the membrane with OPC3 water, we simply changed the water parameters and performed another 500~ps NVT and 10~ns NPT equilibration at 300~K and 1~atm.

\subsection{Ionization} \label{SI:s:ionization}

We reparameterized the TMC fragments with one and two ionized moieties with GAFF and AM1BCC ELF10 for the partial charges. We then performed the same charge reassignment procedure from the polymer termination step. Rather than four constraints, the ionized membrane required seven constraints:

\begin{itemize}
    \item Two terminated MPD fragments had to balance a single linear TMC fragment.
    \item A single crosslinked MPD fragment had to balance two terminated, fully protonated TMC fragments.
    \item Three terminated MPD fragments had to balance one crosslinked TMC fragment.
    \item The charge on a linear, ionized TMC fragment had to equal the charge on a linear, protonated TMC fragment minus one.
    \item The charge on a terminated, fully ionized TMC fragment had to equal the charge on a terminated, protonated TMC fragment minus two.
    \item The charge on a terminated, partially ionized TMC fragment had to equal the charge on a terminated, protonated TMC fragment minus two.
    \item Finally, the total negative charge must be equal to the number of ionized groups in the membrane.
\end{itemize}

\begin{figure}[H]
    \centering
    \includegraphics[width=0.9\textwidth]{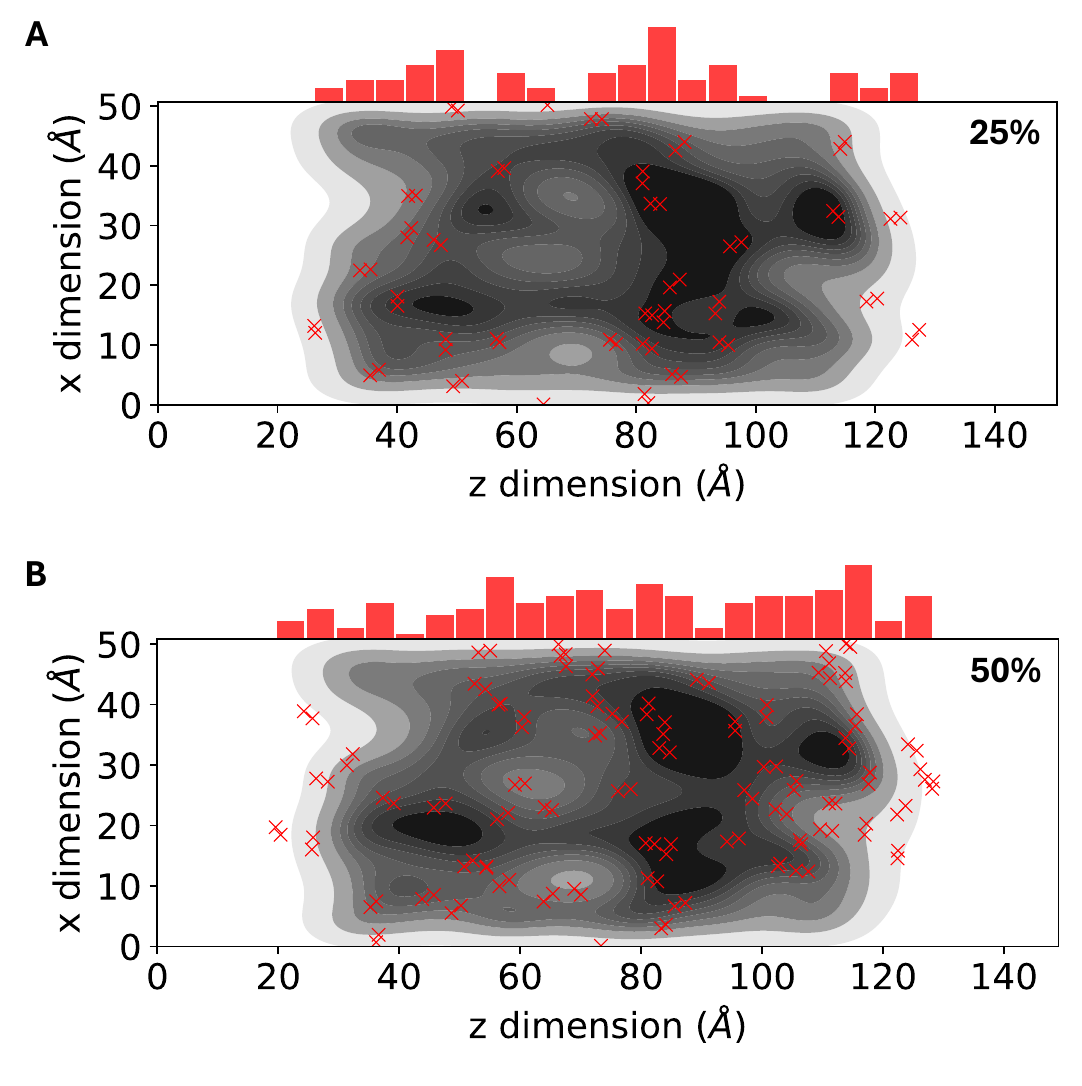}
    \caption{\textbf{Distributions of the ionized moieties across the membrane for the 25\% and 50\% ionized models.} The carboxylate oxygen atoms are shown as red x's. The distribution of the carboxylate oxygens is shown in the marginal axis. These distributions are simply the number within each bin. Each bin has a z-width of 5~\AA. (\textbf{A}) The 25\% ionized membrane shows some slight localization of the carboxylate groups in the densest region of the polymer. (\textbf{B}) The 50\% has more evenly distributed carboxylate groups.}
    \label{SI:fig:membrane_ionized}
\end{figure}

After ionizing and reparameterizing the polymer, we re-equilibrated the charged membrane. Equilibration consisted of the same 500~ps NVT and 10~ns NPT from hydration, in addition to another 100~ns NPT. The production (100~ns) NPT simulation used Parrinello-Rahman pressure coupling (\texttt{tau\_p} = 10~ps) at 1~bar. The final hydrated densities of both the 25\% and 50\% ionized membrane were 1.33~g/cm$^3$. These slightly decreased densities from the charge neutral membranes indicate some swelling, consistent with other simulation studies~\cite{liu_molecular_2022, kolev_molecular_2015}. The x-y dimension for the 25\% ionized membrane increased from 5.03~nm to 5.13~nm, and for the 50\% ionized membrane, it increased to 5.19~nm. In Figure~\ref{SI:fig:membrane_ionized}, we show the distribution of the carboxylate groups for the 25\% and 50\% ionized membranes. The red x's indicate the oxygen atoms in the carboxylate groups. Notably, both membranes include significant numbers of interfacial and confined fixed charges.

\section{Validation of methods}

\subsection{Hydration}

\begin{figure}[H]
    \centering
    \includegraphics[width=0.75\textwidth]{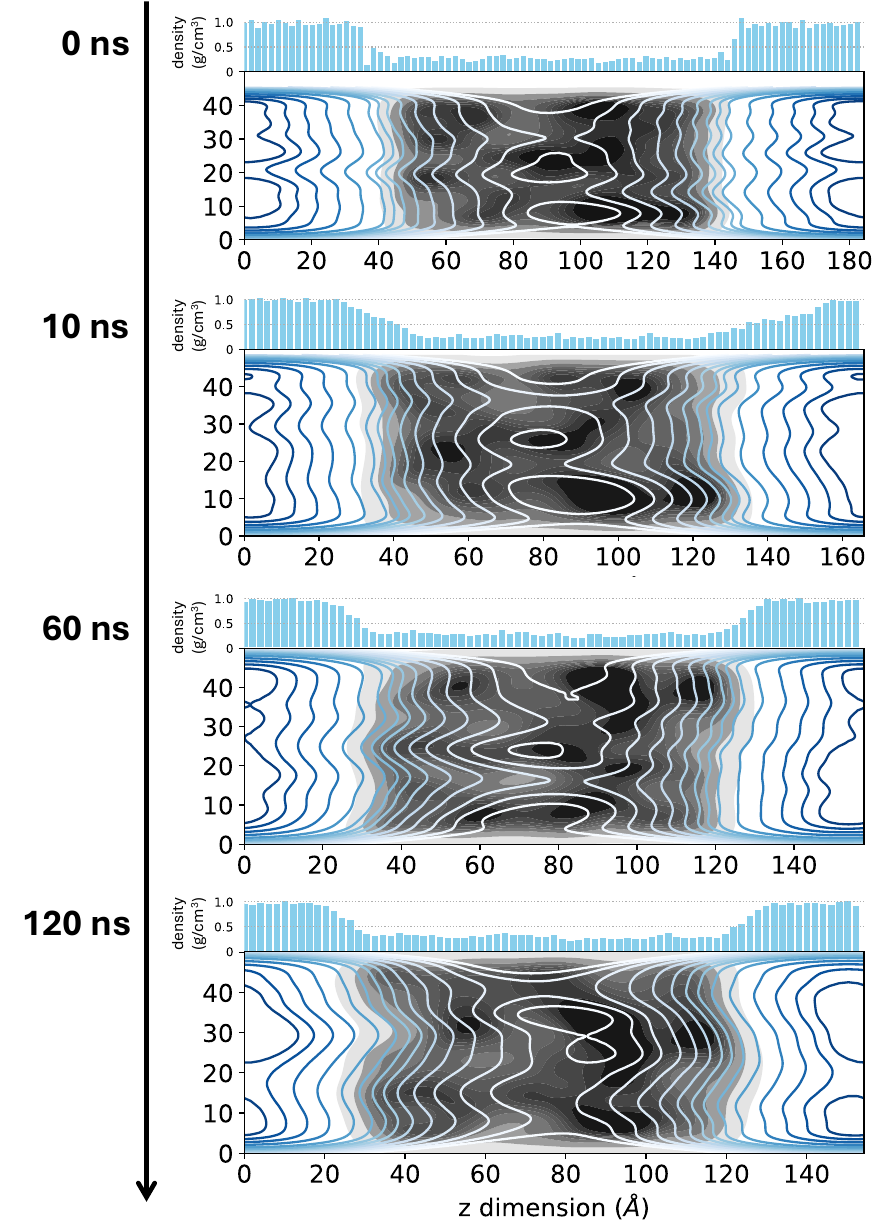}
    \caption{\textbf{Iterative hydration process.} Snapshots of the membrane were taken after NPT steps during hydration with TIP3P water. 0~ns shows the distribution of water that was preinserted into the membrane and added to the reservoir. Each snapshot is plotted as described in Figure~\ref{fig:membrane_hydration}B.}
    \label{SI:fig:hydration_process}
\end{figure}

\begin{figure}[H]
    \centering
    \includegraphics[width=0.75\textwidth]{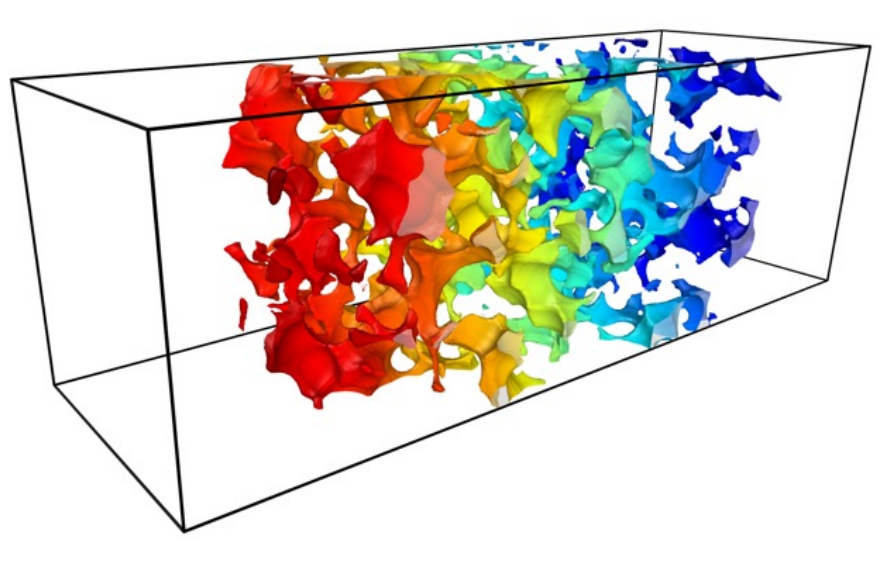}
    \caption{\textbf{Snapshot of the water network in the membrane.} The water surface mesh is constructed with the Gaussian density method as implemented in Ovito version 3.8.4~\cite{stukowski_visualization_2010} (resolution = 600, radius scaling = 100\%, isovalue = 0.05. The surface is colored according to its z-coordinate.}
    \label{SI:fig:percolating_water}
\end{figure}

We demonstrate that the membrane comes to an equilibrium state when water molecules diffuse into membrane nanovoids. We built the polyamide membrane with a natural interface by imposing harmonic walls to the z boundaries. We then put this interface in contact with a water reservoir as shown in Figure~\ref{SI:fig:hydration_process} at 0~ns. Water then diffused into the membrane, indicated by the linear water density through the interface (Figure~\ref{SI:fig:hydration_process} at 10~ns and 60~ns). Notably, the size of the water reservoir significantly shrank during the hydration process, decreasing from approximately 8~nm thick to approximately 5~nm thick. The minimum water density bin goes from 0.14~g/cm$^3$ to 0.21~g/cm$^3$ after the 120~ns hydration process, indicating significant diffusion throughout the membrane. Furthermore, the average water density in the middle 50\% of the membrane goes from 0.27~g/cm$^3$ after preinsertion to 0.31~g/cm$^3$ after hydration. The contours in Figure~\ref{SI:fig:hydration_process} show that water moves deeper into the membrane and prefers the regions of low polymer density. We do not perform a rigorous percolation analysis of the water channels, but our results do not provide strong evidence that water channels percolate across the entire z-dimension of the membrane, as shown in Figure~\ref{SI:fig:percolating_water}.

\subsection{Ion equilibration into the membrane}

\begin{figure}[H]
    \centering
    \includegraphics[width=0.475\textwidth]{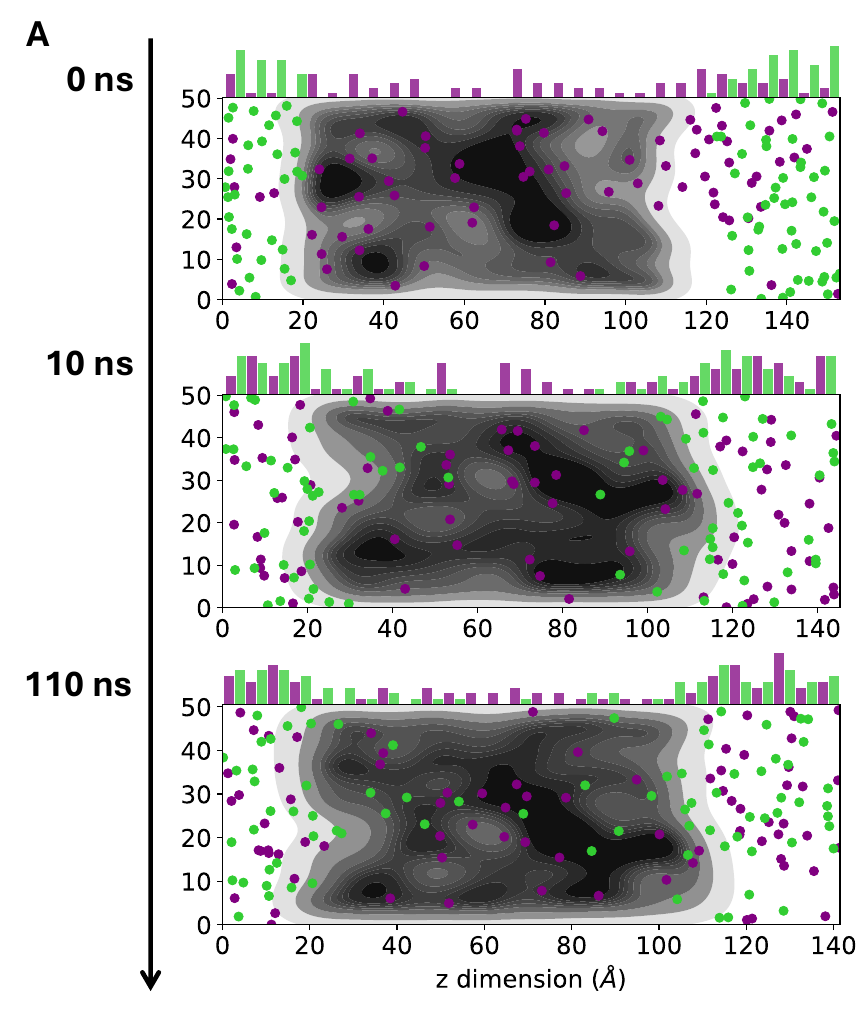}
    \includegraphics[width=0.452\textwidth]{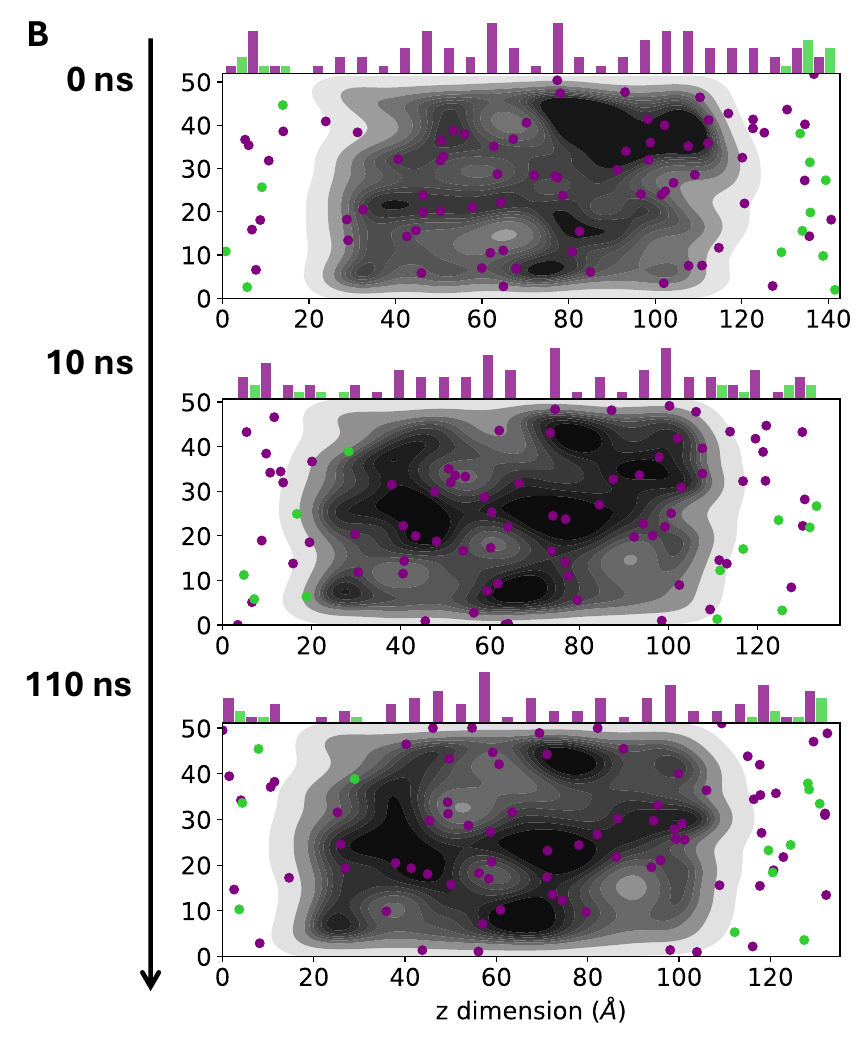}
    \caption{\textbf{Ion equilibration process for the 0\% ionized membrane and the 50\% ionized membrane.} (\textbf{A}) Snapshots of the ion equilibration process for the 0\% ionized membrane soaked in a NaCl solution. This simulation contains 79 Na$^+$ and 79 Cl$^-$ total. The Na$^+$ ions are purple, and the Cl$^-$ ions are green. The number density distributions for the Na$^+$ and Cl$^-$ ions are shown in the marginal axis. Each bin has a z-width of 5~\AA. }
    \label{SI:fig:ion_equilibration}
\end{figure}

After inserting ions into the membrane and into the water reservoir, we tracked the ion equilibration process to determine where ions will typically be during operation. For the 0\% ionized membranes (e.g. Figure~\ref{SI:fig:ion_equilibration}A), we added the same number of cations into the membrane that would be needed to balance the charge of a 50\% ionized membrane in order to efficiently sample ion environments within the membrane. However, this procedure introduces a net positive charge into the membrane. As a result, anions partition from the solution into the membrane, even within the relatively short equilibration steps. We observe anions that penetrated fully into the membrane during the 110~ns equilibration procedure as shown in Figure~\ref{SI:fig:ion_equilibration}A. On the other hand, the ionized membranes (e.g. 50\% in Figure~\ref{SI:fig:ion_equilibration}B) do not see any significant anion penetration. We inserted positive counterions nearby randomly selected carboxylate groups throughout these ionized membranes, such that there is not charge heterogeneity across the membrane. We carefully detail the resulting coordination environments for the cations within the membrane in Section~\nameref{s:results}.

\begin{figure}[H]
    \centering
    \includegraphics[width=0.75\textwidth]{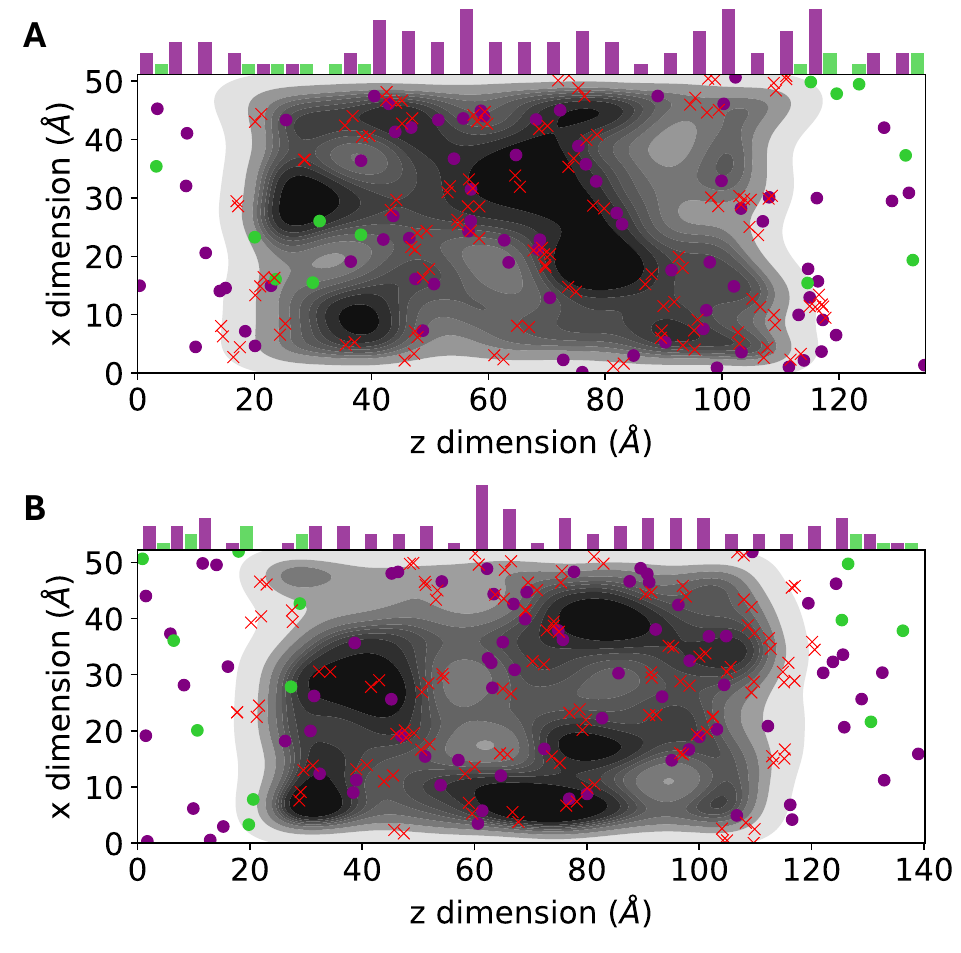}
    \caption{\textbf{Snapshots after 2~$\mu$s simulations for different insertion methods.} (\textbf{A}) Na$^+$ were inserted near carboxylate groups. (\textbf{B}) Na$^+$ were inserted randomly into the membrane. Plots are generated as described in Figures~\ref{SI:fig:membrane_ionized} and~\ref{SI:fig:ion_equilibration} for a 50\% ionized membrane.}
    \label{SI:fig:ion_insertion_random}
\end{figure}

We hypothesized that within an ionized membrane, the equilibrium configuration most likely corresponds to cations coordinated with the ionized carboxylate groups~\cite{ritt_ionization_2020, chen_facile_2017}. Accordingly, we inserted cations nearby carboxylate groups. To test this hypothesis, we also performed ion equilibration and long time-scale production simulations with Na$^+$ ions inserted randomly in the membrane. Figure~\ref{SI:fig:ion_insertion_random} shows snapshots from the end of the 2~$\mu$s simulations for ions placed near the carboxylates (A) and placed randomly (B). The distributions of Na$^+$ are structured similarly, with peaks near z~=~55~\AA~and z~=~100~\AA. The peak near 55~\AA~corresponds to a large concentration of carboxylate groups (indicated with red x's), and the peak near 100~\AA~corresponds to the interface with a significant number of carboxylate groups. These observations qualitatively support our hypothesis.

\subsection{Pore size distributions}

\begin{figure}[H]
    \centering
    \includegraphics[width=0.5\textwidth]{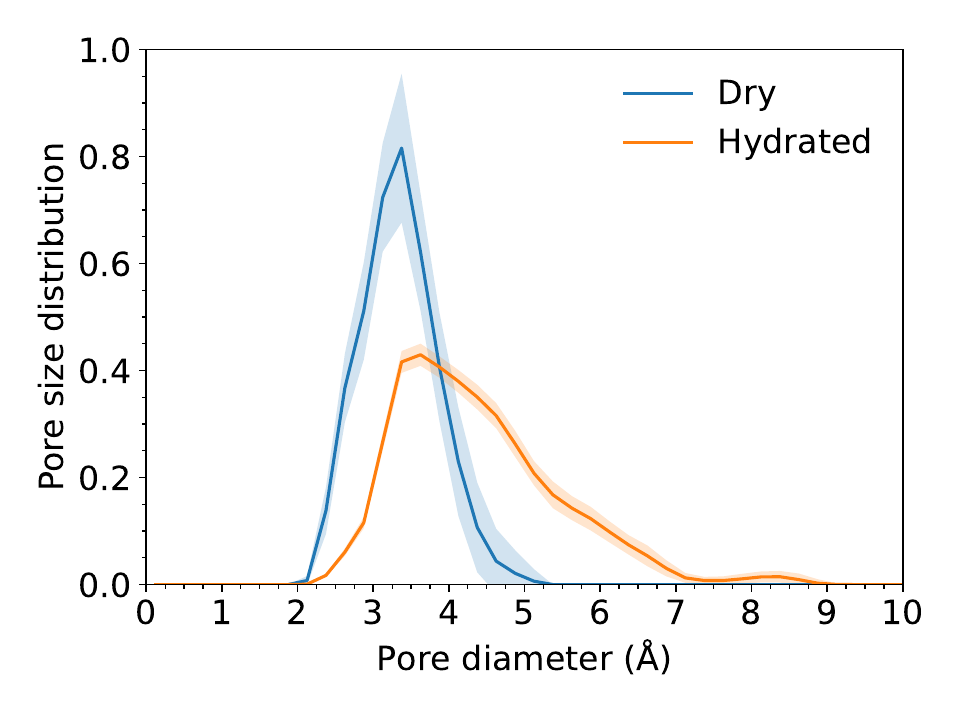}
    \caption{\textbf{Pore size distribution for the dry and hydrated membrane.} The error shown is the standard deviation over the 100 frames that were used to calculate the pore size distribution.}
    \label{SI:fig:pore_size_distributions}
\end{figure}

To further characterize the membrane structure, we estimated the pore size distribution for both the equilibrated dry membrane and the fully hydrated membrane. We used PoreBlazer version 4.0~\cite{sarkisov_materials_2020} to calculate the pore size distribution. PoreBlazer estimates the pore size by a geometric approach. It identifies spheres with the largest possible diameter without overlapping the atomic structure of the membrane. We performed time averages over 100 frames from the end of the 21-step equilibration and from the end of the hydration process. The resulting distributions are shown in Figure~\ref{SI:fig:pore_size_distributions}.

Our pore size distributions show primary pore diameters of about 3~\AA~and 4~\AA~for the dry and hydrated membranes, respectively. These pore sizes are comparable to experimental PALS and small-angle X-ray scattering, which report a range of 4.4-5.0~\AA~for hydrated polyamide membranes~\cite{lee_water_2013, singh_synchrotron_2012}. While our simulations do not show both the commonly reported ``network" (4~\AA) and ``aggregate" (8~\AA) pores~\cite{kim_positron_2005, sharma_dynamics_2009}, our unimodal distributions are consistent with other simulation studies~\cite{liyana-arachchi_ultrathin_2016, vickers_molecular_2022, ding_molecular_2014}.

\subsection{Ion insertion methods} \label{SI:s:ion_insertion_validation}

\begin{figure}[H]
    \centering
    \includegraphics[width=0.75\textwidth]{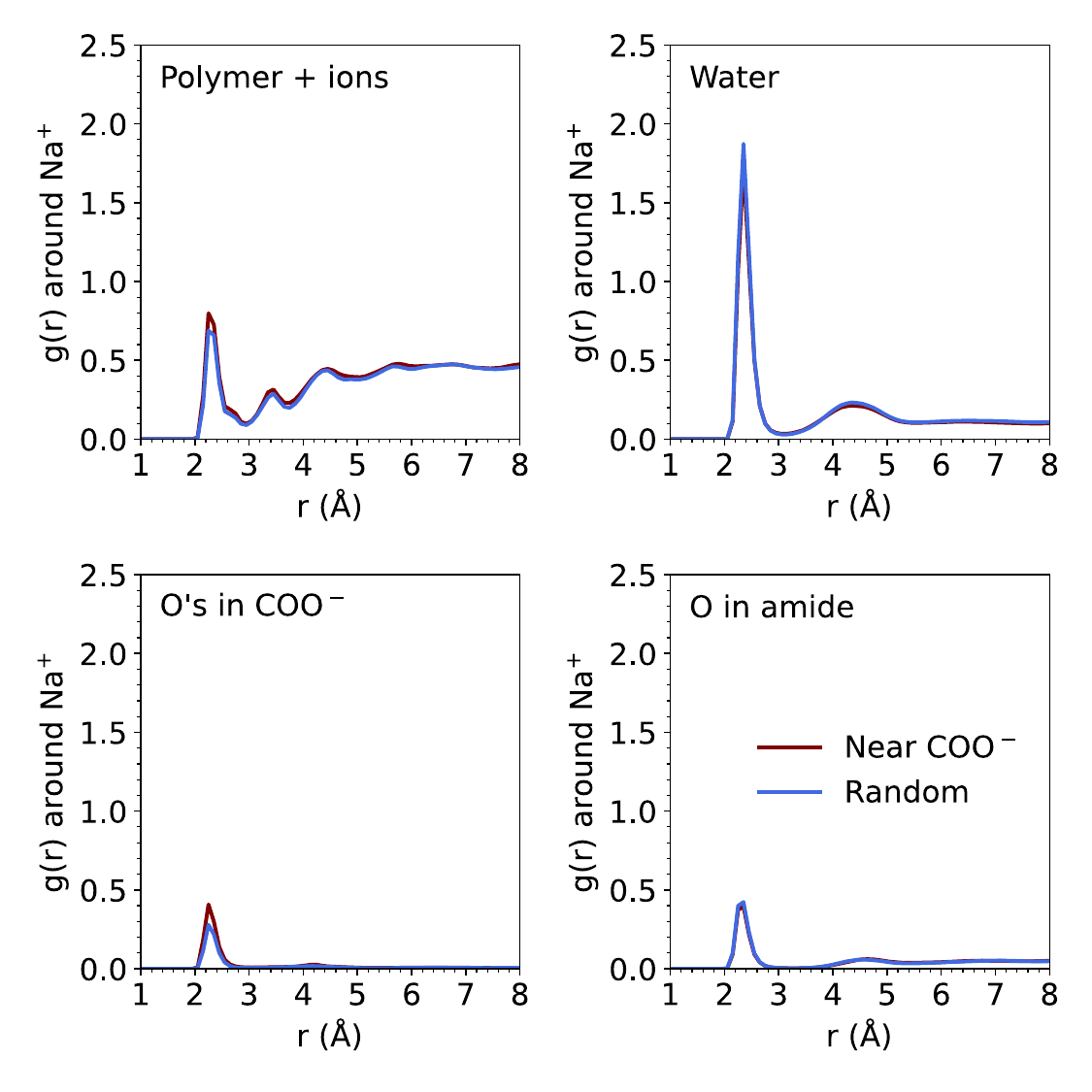}
    \caption{\textbf{RDFs comparing Na$^+$ insertion methods.} RDFs were generated between Na$^+$ and the specified groups within the inner 50\% of the membrane. ``Near COO$^-$" indicates simulations where ions were started near carboxylate groups, and ``Random" refers to simulations where ions were inserted randomly. In the ``Polymer + ions'' and ``Water'' RDFs, contributions from hydrogen atoms are not included.}
    \label{fig:rdf_Na_insertion}
\end{figure}

\begin{figure}[H]
    \centering
    \includegraphics[width=0.75\textwidth]{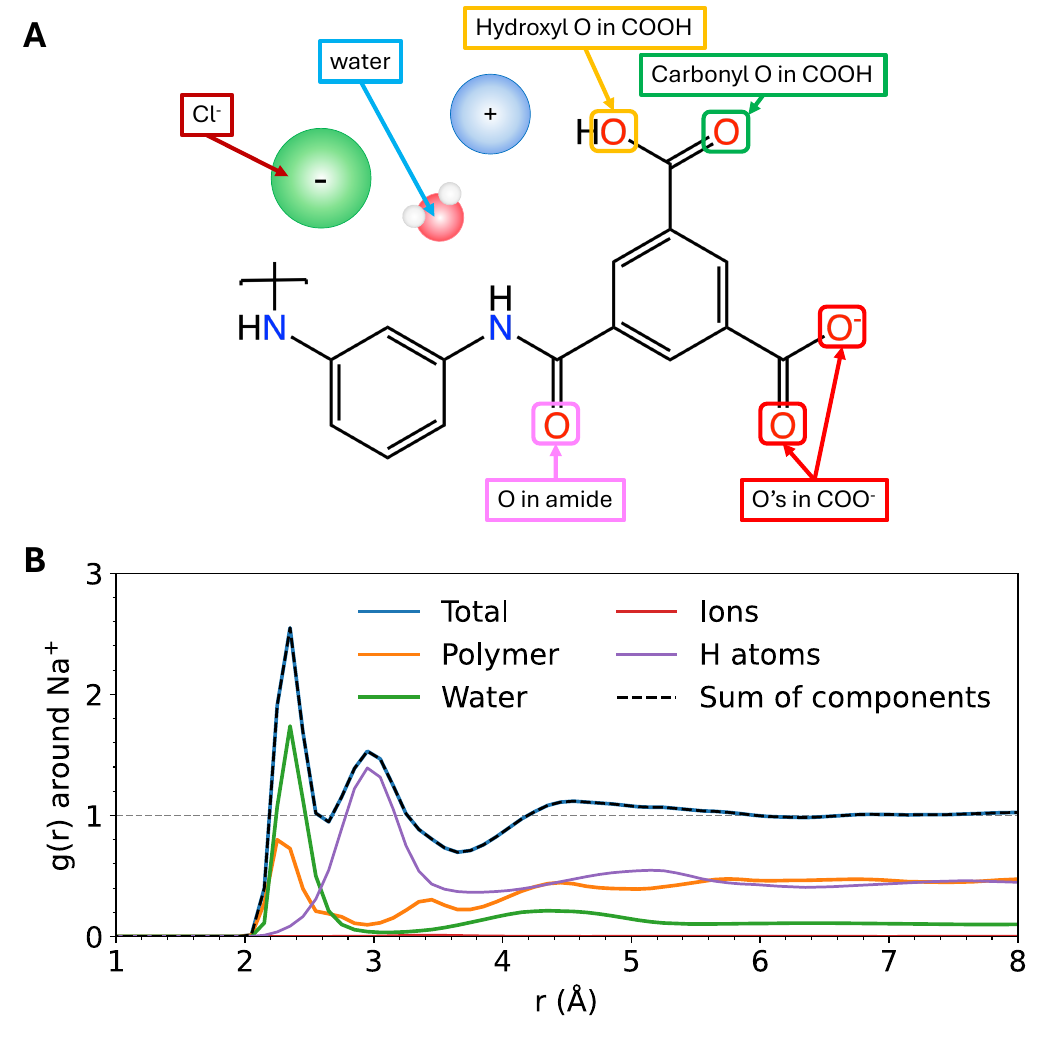}
    \caption{\textbf{Chemical structure of the species in simulation and illustration of renormalization used in Supporting Information Section~\nameref{SI:s:ion_insertion_validation}} We highlight the groups considered for the chemical breakdown of RDFs, coordination shells, and nearest neighbors.}
    \label{SI:fig:rdf_selections}
\end{figure}

We return to the hypothesis that the equilibrium configuration involves ions coordinating with ionized groups in the membrane. If cations bind tightly to the fixed charges, they will move towards the fixed charges when placed in far-from-equilibrium starting configurations. Figure~\ref{fig:rdf_Na_insertion} shows the RDFs for Na$^+$ ions in the simulations where cations were placed near carboxylate groups (maroon) and placed randomly in the membrane (royal blue). We depict the selections here and others we considered in Figure~\ref{SI:fig:rdf_selections}. Notably, neither the ``Polymer + ions'' selection nor the ``Water'' selection include contributions from hydrogen atoms. Contributions from other ions in the membrane are included in the ``Polymer + ions'' RDFs, but their contributions are negligible. We find that the cation-O peak for the ionized carboxylate groups (COO$^-$) is more prominent ($\approx$1.5x) in the simulations where cations were started near those groups. Since the COO$^-$ density is higher, the water density is lower, and the coordination peak in the ``Water'' panel decreases. When the ion is not coordinated with a COO$^-$ oxygen, it fills the coordination shell with a water molecule. However, all other coordination structure is unchanged, which supports our hypothesis that cations will coordinate with COO$^-$ at equilibrium. The second most polar group in the polymer, after the COO$^-$, is the carboxyl group in the amide crosslinking bond. The cation-O peak for this carboxyl group does not change between the insertion cases, which indicates that Na$^+$ does not coordinate more strongly with the polymer when far from COO$^-$ groups. Rather it is more hydrated.

The randomly placed ions appear to approach the same structure as those placed near COO$^-$, but the slight difference may indicate that the simulation with randomly inserted ions has not fully equilibrated after the 2~$\mu$s. The highly crosslinked RO membrane significantly hinders ion transport, so if a cation was placed very far from equilibrium during random insertion, the necessary equilibration times may be beyond what is reasonably achievable in molecular simulation.

\section{Selections for the RDFs}

\section{Ion-oxygen distances}

\begin{table}[H]
\centering
\caption{\textbf{Ion-oxygen distances in solution.} The first oxygen atom neighbor distances are shown with their standard deviations as uncertainty.}
\begin{tabular}{|l|l|l|l|l|l|}
\hline
\textbf{Ion}       & \textbf{Oxygen} & \textbf{1$^{\text{st}}$ Peak (Å)} & \textbf{Peak Width (Å)} & \textbf{Avg. to Cutoff (Å)} & \textbf{1$^{\text{st}}$ Neighbor (Å)} \\ \hline
Na$^+$    & Water  & 2.365          & 0.251          & 2.437 $\pm$ 0.170       & 2.275 $\pm$ 0.053    \\ \hline
K$^+$     & Water  & 2.715          & 0.309          & 2.846 $\pm$ 0.206       & 2.620 $\pm$ 0.066    \\ \hline
Rb$^+$    & Water  & 2.855          & 0.323          & 3.001 $\pm$ 0.214       & 2.751 $\pm$ 0.071    \\ \hline
Ca$^{2+}$ & Water  & 2.335          & 0.180          & 2.370 $\pm$ 0.099       & 2.260 $\pm$ 0.037    \\ \hline
Sr$^{2+}$ & Water  & 2.535          & 0.196          & 2.566 $\pm$ 0.099       & 2.444 $\pm$ 0.038    \\ \hline
\end{tabular}
\label{tab:ion_oxygen_solution}
\end{table}

\begin{table}[H]
\centering
\caption{\textbf{Ion-oxygen distances in 50\% ionized membrane.} Distances were reported in the same way as Table~\ref{tab:ion_oxygen_solution}.}
\begin{tabular}{|l|l|r|r|r|r|}
\hline
\textbf{Ion}               & \textbf{Oxygen} & \multicolumn{1}{l|}{\textbf{1$^{\text{st}}$ Peak (Å)}} & \multicolumn{1}{l|}{\textbf{Peak Width (Å)}} & \multicolumn{1}{l|}{\textbf{Avg. to Cutoff (Å)}} & \multicolumn{1}{l|}{\textbf{1$^{\text{st}}$ Neighbor (Å)}} \\ \hline
\multirow{4}{*}{Na$^+$}    & All             & 2.325                                                  & 0.253                                        & 2.399 $\pm$ 0.161                                & \multirow{4}{*}{2.240 $\pm$ 0.057}                         \\ \cline{2-5}
                           & Water           & 2.335                                                  & 0.242                                        & 2.419 $\pm$ 0.156                                &                                                            \\ \cline{2-5}
                           & Amide           & 2.295                                                  & 0.243                                        & 2.381 $\pm$ 0.162                                &                                                            \\ \cline{2-5}
                           & COO$^-$         & 2.265                                                  & 0.232                                        & 2.330 $\pm$ 0.141                                &                                                            \\ \hline
\multirow{4}{*}{K$^+$}     & All             & 2.685                                                  & 0.309                                        & 2.805 $\pm$ 0.199                                & \multirow{4}{*}{2.590 $\pm$ 0.068}                         \\ \cline{2-5}
                           & Water           & 2.715                                                  & 0.299                                        & 2.827 $\pm$ 0.200                                &                                                            \\ \cline{2-5}
                           & Amide           & 2.665                                                  & 0.296                                        & 2.762 $\pm$ 0.203                                &                                                            \\ \cline{2-5}
                           & COO$^-$         & 2.655                                                  & 0.303                                        & 2.782 $\pm$ 0.221                                &                                                            \\ \hline
\multirow{4}{*}{Rb$^+$}    & All             & 2.845                                                  & 0.329                                        & 2.985 $\pm$ 0.247                                & \multirow{4}{*}{2.715 $\pm$ 0.073}                         \\ \cline{2-5}
                           & Water           & 2.825                                                  & 0.321                                        & 3.008 $\pm$ 0.230                                &                                                            \\ \cline{2-5}
                           & Amide           & 2.805                                                  & 0.330                                        & 2.926 $\pm$ 0.248                                &                                                            \\ \cline{2-5}
                           & COO$^-$         & 2.795                                                  & 0.327                                        & 2.936 $\pm$ 0.243                                &                                                            \\ \hline
\multirow{4}{*}{Ca$^{2+}$} & All             & 2.315                                                  & 0.180                                        & 2.326 $\pm$ 0.090                                & \multirow{4}{*}{2.216 $\pm$ 0.042}                         \\ \cline{2-5}
                           & Water           & 2.315                                                  & 0.167                                        & 2.345 $\pm$ 0.084                                &                                                            \\ \cline{2-5}
                           & Amide           & 2.315                                                  & 0.174                                        & 2.340 $\pm$ 0.108                                &                                                            \\ \cline{2-5}
                           & COO$^-$         & 2.245                                                  & 0.171                                        & 2.281 $\pm$ 0.086                                &                                                            \\ \hline
\multirow{4}{*}{Sr$^{2+}$} & All             & 2.515                                                  & 0.204                                        & 2.539 $\pm$ 0.104                                & \multirow{4}{*}{2.408 $\pm$ 0.047}                         \\ \cline{2-5}
                           & Water           & 2.525                                                  & 0.190                                        & 2.555 $\pm$ 0.095                                &                                                            \\ \cline{2-5}
                           & Amide           & 2.505                                                  & 0.219                                        & 2.554 $\pm$ 0.119                                &                                                            \\ \cline{2-5}
                           & COO$^-$         & 2.445                                                  & 0.184                                        & 2.474 $\pm$ 0.098                                &                                                            \\ \hline
\end{tabular}
\label{tab:ion_oxygen_polymer}
\end{table}

\section{Distributions of the nearest neighbor distances for Na$^+$, Rb$^+$, and Sr$^{2+}$} \label{SI:s:distance_distributions}

 \begin{figure}[H]
    \centering
    \includegraphics[width=0.75\textwidth]{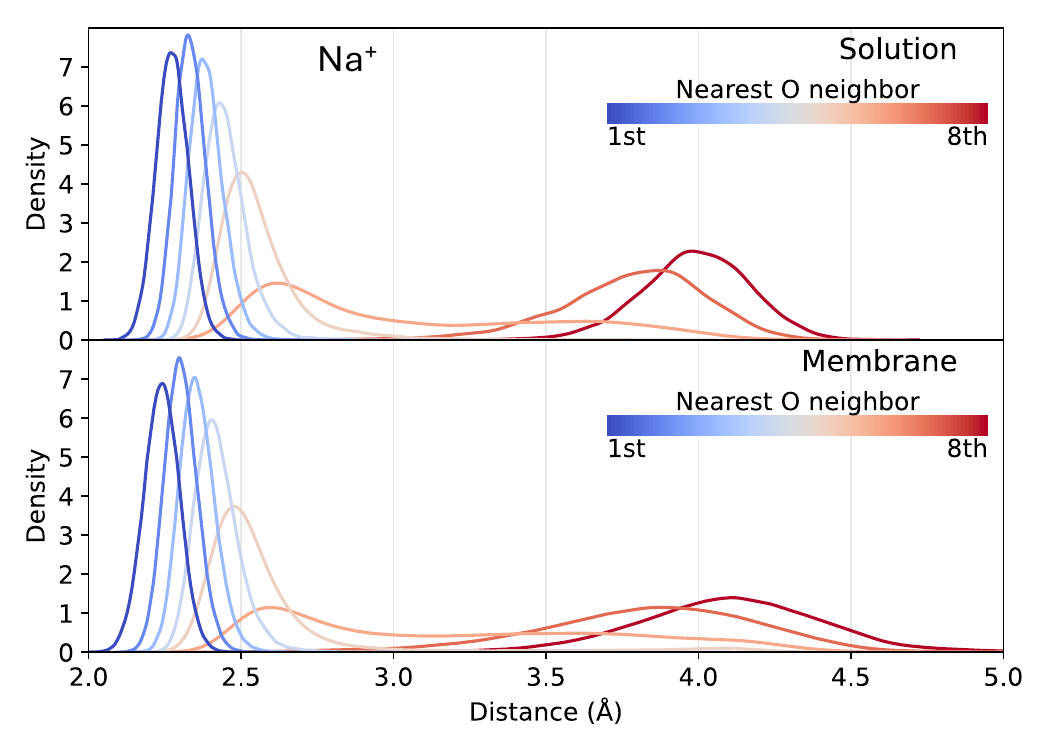}
    \caption{\textbf{Distribution of nearest oxygen neighbor distances for Na$^+$}} 
    \label{SI:fig:neighbor_distances_Na}
\end{figure}

 \begin{figure}[H]
    \centering
    \includegraphics[width=0.75\textwidth]{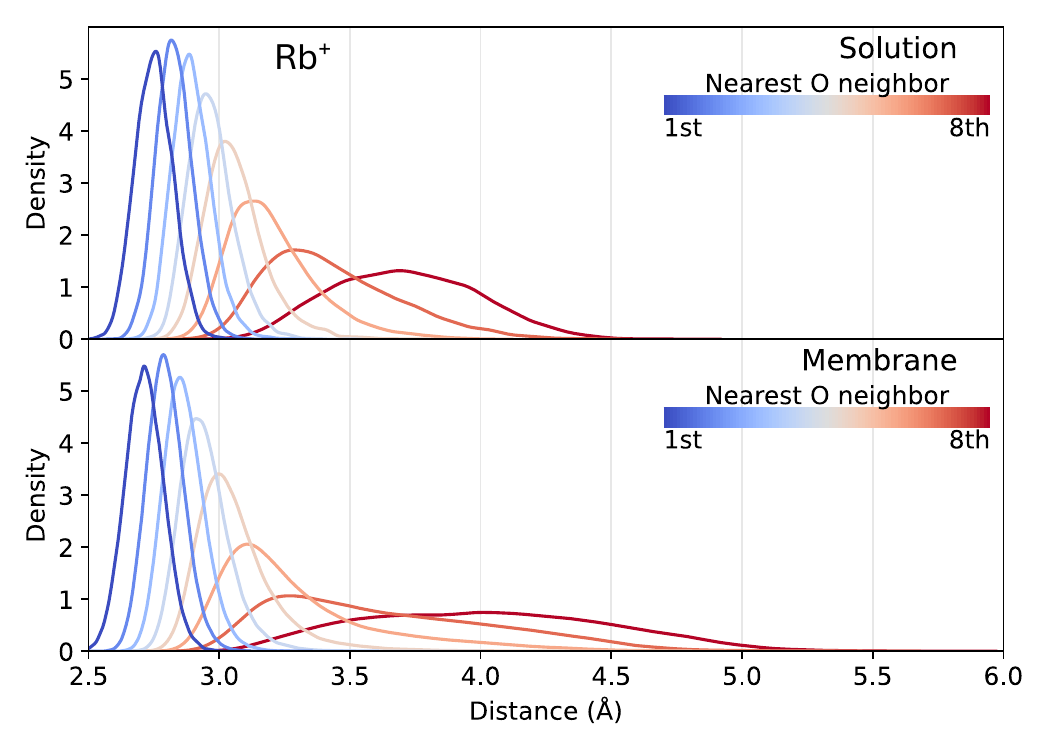}
    \caption{\textbf{Distribution of nearest oxygen neighbor distances for Rb$^+$}} 
    \label{SI:fig:neighbor_distances_Rb}
\end{figure}

 \begin{figure}[H]
    \centering
    \includegraphics[width=0.75\textwidth]{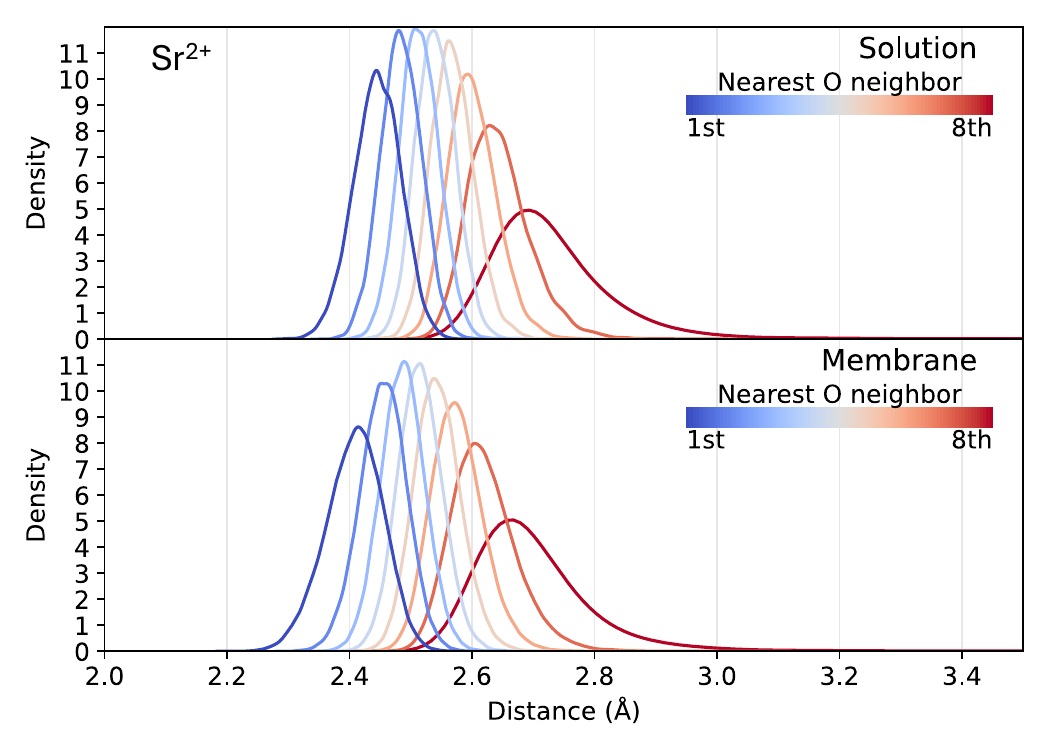}
    \caption{\textbf{Distribution of nearest oxygen neighbor distances for Sr$^{2+}$}} 
    \label{SI:fig:neighbor_distances_Sr}
\end{figure}

\section{Average coordination numbers in solution}

 \begin{figure}[H]
    \centering
    \includegraphics[width=0.75\textwidth]{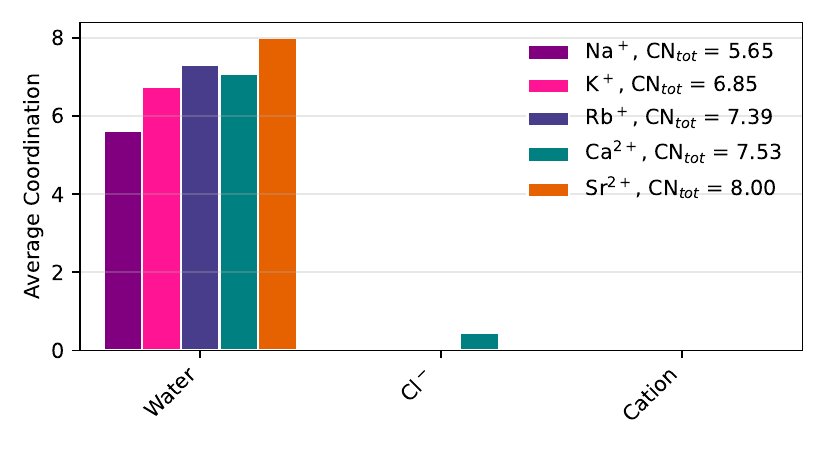}
    \caption{\textbf{Distribution of average coordination number across species in solution.} The average coordination number for each ion is included in the legend.}
    \label{SI:fig:average_coordination_solution}
\end{figure}

\section{Additional RDFs}

In this section, we include all RDFs for both solution and in polymer. We plot the coordination shell cutoff for each ion-species interaction as a red dashed line. 

\subsection{Solution RDFs}

 \begin{figure}[H]
    \centering
    \includegraphics[width=\textwidth]{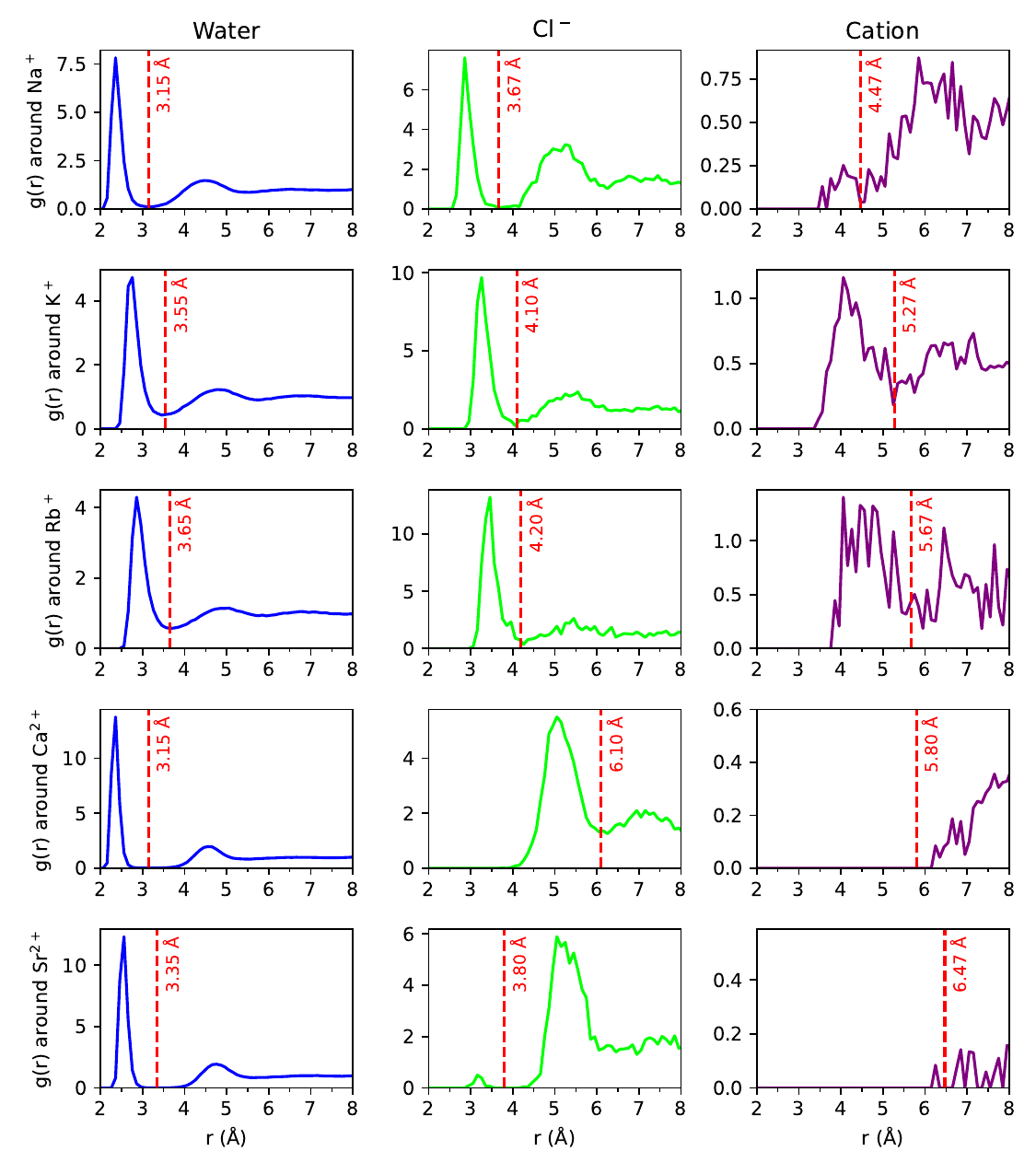}
    \caption{\textbf{RDFs for all ions in solution.} We plot the cation-water, cation-anion, and cation-cation RDFs for Na$^+$, K$^+$, Rb$^+$, Ca$^{2+}$, and Sr$^{2+}$. Coordination shell cutoffs are red dashed lines.}
    \label{SI:fig:solution_RDFs}
\end{figure}

\subsection{Membrane RDFs} \label{SI:s:membrane_RDFs}

 \begin{figure}[H]
    \centering
    \includegraphics[width=\textwidth]{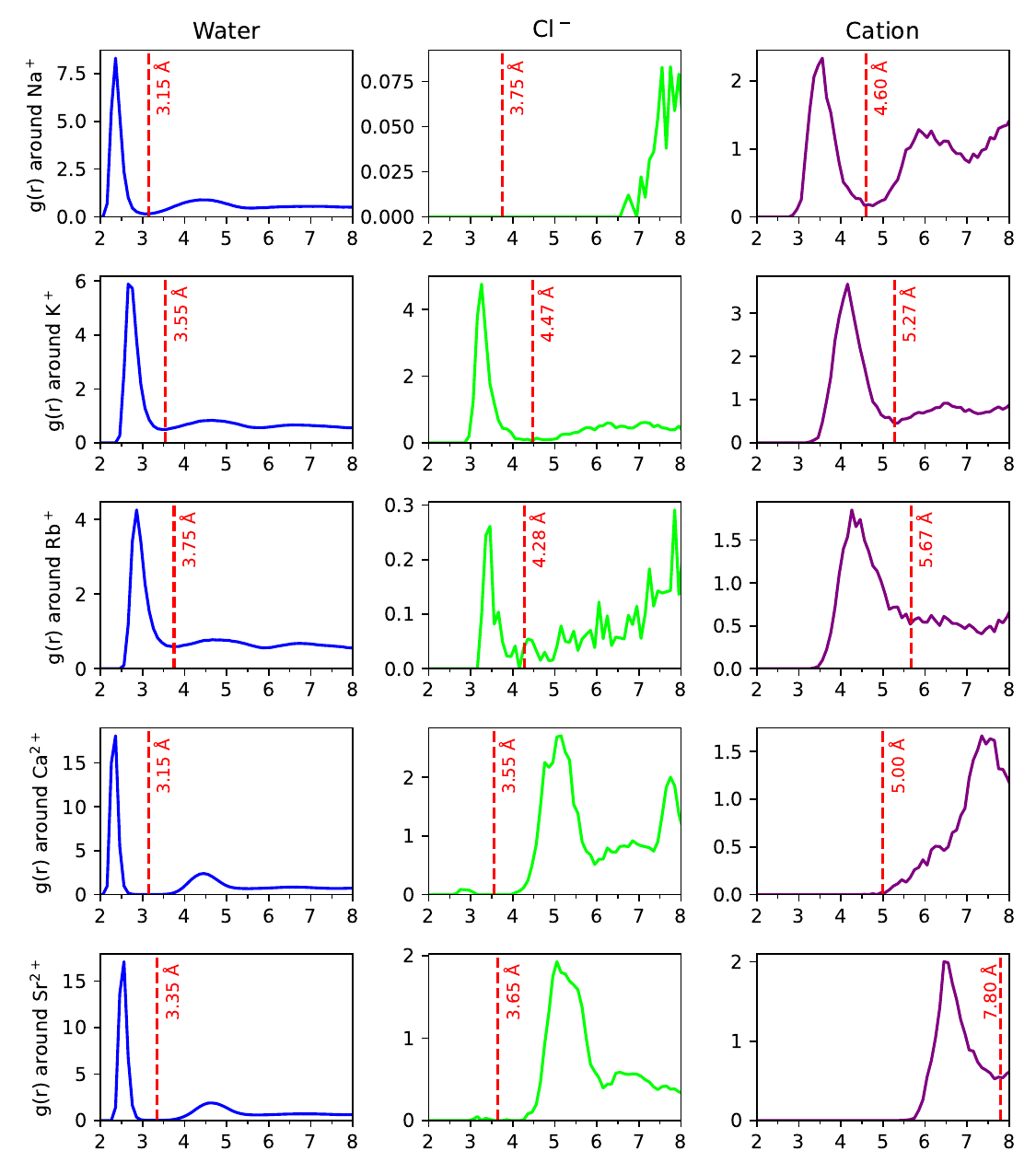}
    \caption{\textbf{RDFs for all ions in the 50\% ionized membrane.} We plot the cation-water, cation-anion, and cation-cation RDFs for Na$^+$, K$^+$, Rb$^+$, Ca$^{2+}$, and Sr$^{2+}$. Coordination shell cutoffs are red dashed lines.}
    \label{SI:fig:membrane_RDFs_1}
\end{figure}

 \begin{figure}[H]
    \centering
    \includegraphics[width=\textwidth]{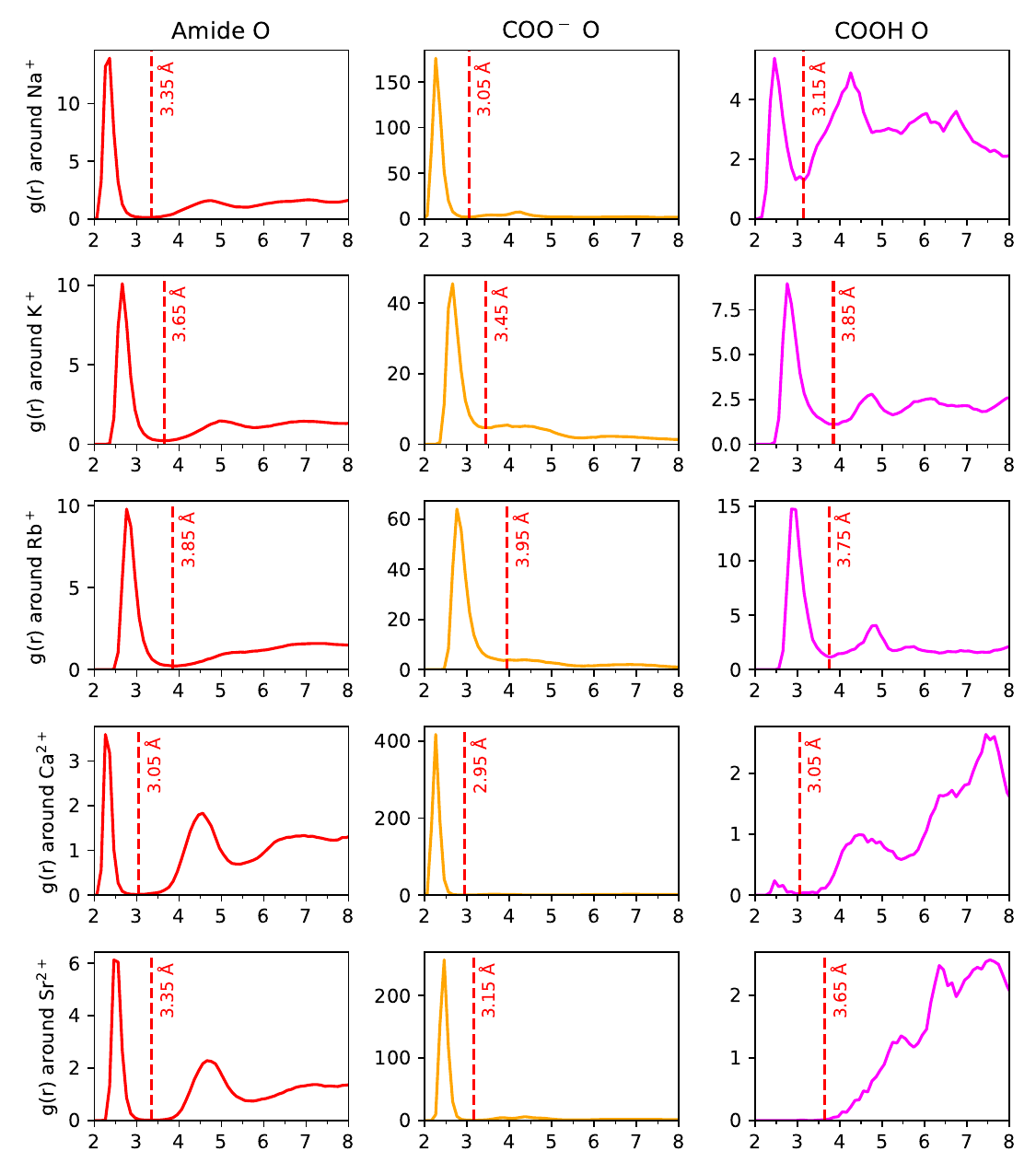}
    \caption{\textbf{RDFs for all ions in the 50\% ionized membrane.} We plot the cation-amide oxygen, cation-carboxylate oxygen, and cation-carboxylic acid oxygen RDFs for Na$^+$, K$^+$, Rb$^+$, Ca$^{2+}$, and Sr$^{2+}$. Coordination shell cutoffs are red dashed lines.}
    \label{SI:fig:membrane_RDFs_2}
\end{figure}

 \begin{figure}[H]
    \centering
    \includegraphics[width=0.333\textwidth]{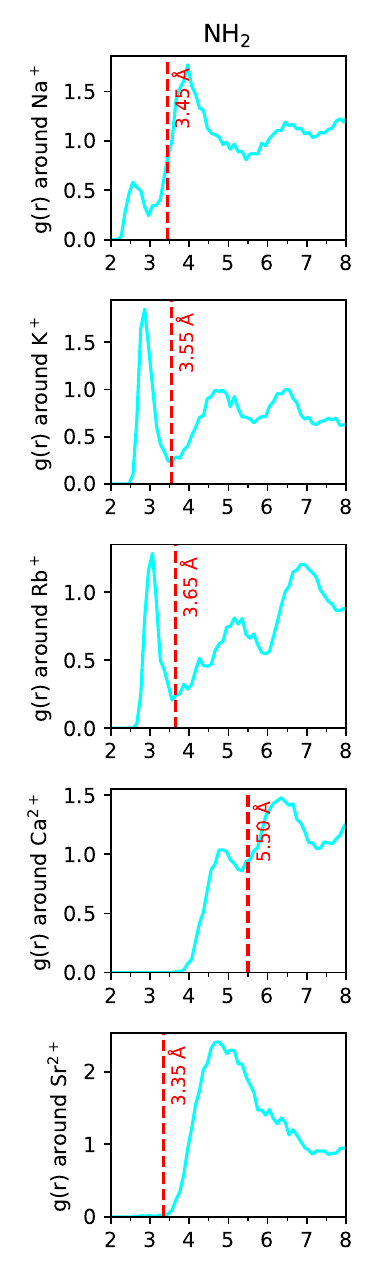}
    \caption{\textbf{RDFs for all ions in the 50\% ionized membrane.} We plot the cation-amine RDFs for Na$^+$, K$^+$, Rb$^+$, Ca$^{2+}$, and Sr$^{2+}$. Coordination shell cutoffs are red dashed lines.}
    \label{SI:fig:membrane_RDFs_3}
\end{figure}

 \begin{figure}[H]
    \centering
    \includegraphics[width=\textwidth]{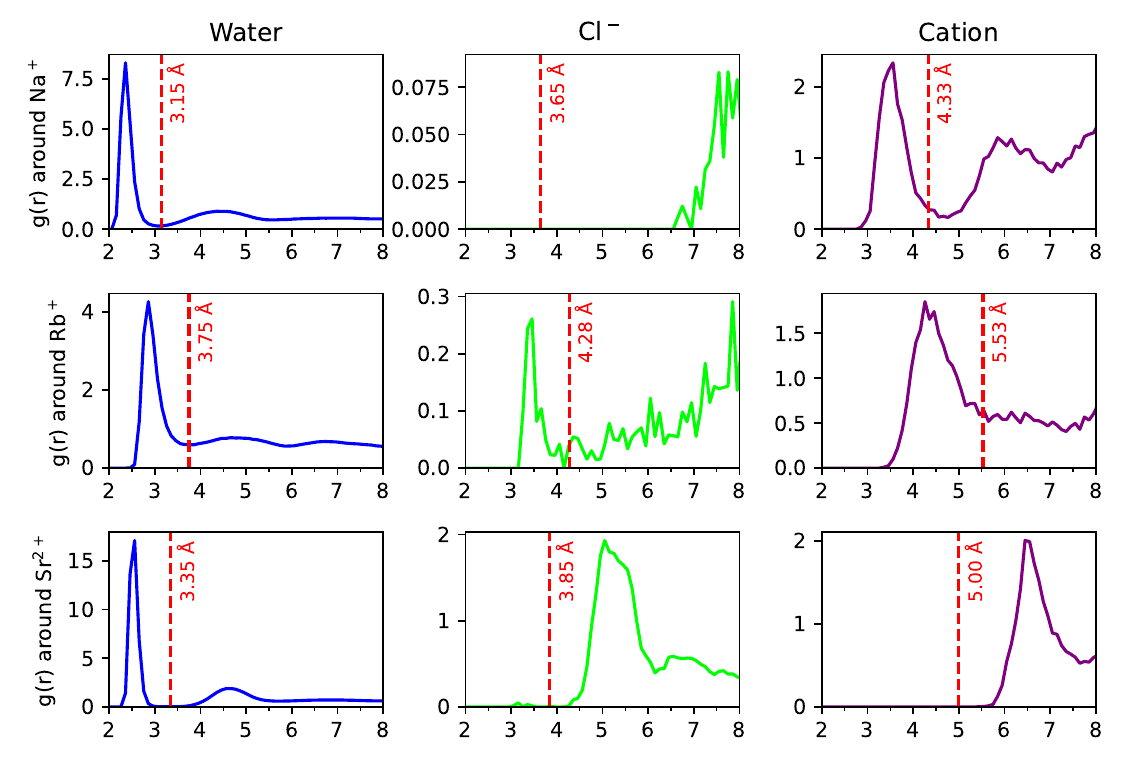}
    \caption{\textbf{RDFs for Na$^+$, Rb$^+$, and Sr$^{2+}$ in the 0\% ionized membrane.} We plot the cation-water, cation-anion, and cation-cation RDFs. Coordination shell cutoffs are red dashed lines.}
    \label{SI:fig:membrane_RDFs_Opct_1}
\end{figure}

 \begin{figure}[H]
    \centering
    \includegraphics[width=\textwidth]{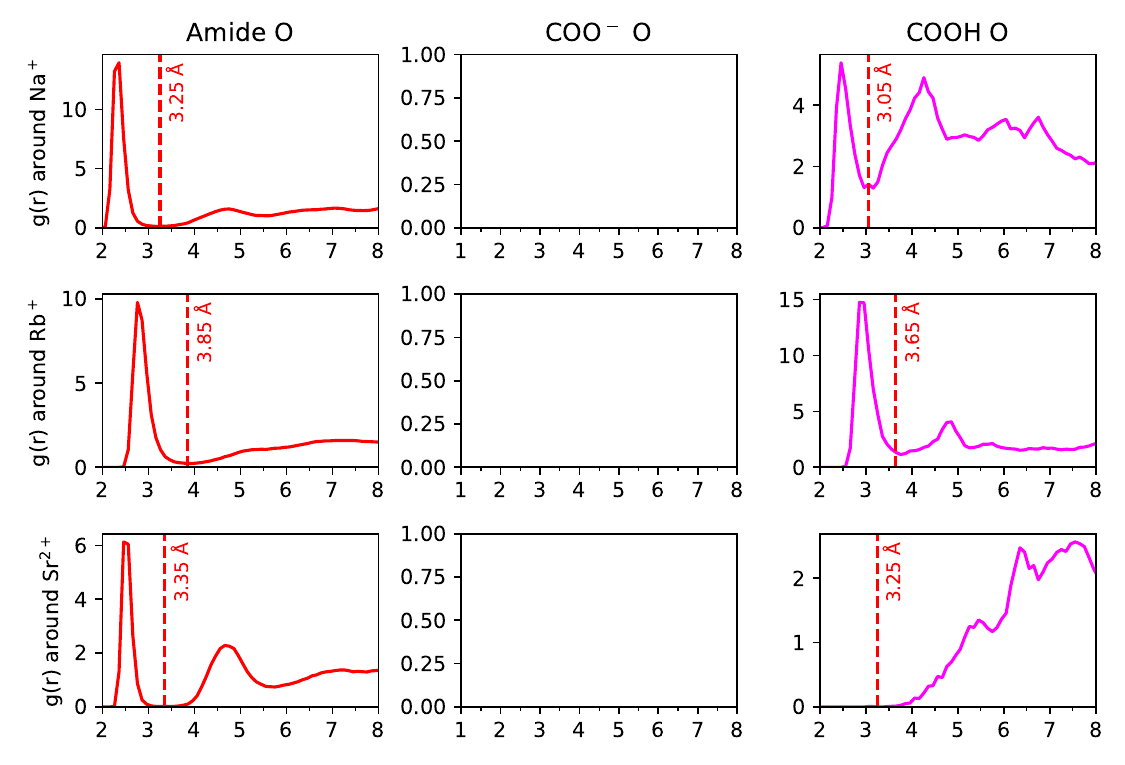}
    \caption{\textbf{RDFs for Na$^+$, Rb$^+$, and Sr$^{2+}$ in the 0\% ionized membrane.} We plot the cation-amide oxygen, cation-carboxylate oxygen, and cation-carboxylic acid oxygen RDFs. Note that no carboxylates are present in the 0\% ionized membrane. Coordination shell cutoffs are red dashed lines.}
    \label{SI:fig:membrane_RDFs_Opct_2}
\end{figure}

 \begin{figure}[H]
    \centering
    \includegraphics[width=0.333\textwidth]{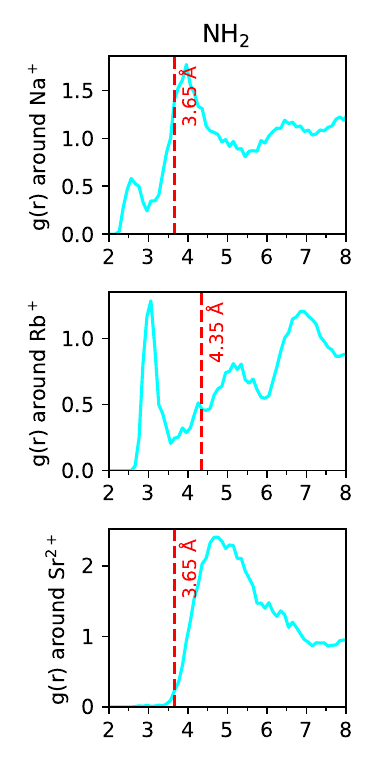}
    \caption{\textbf{RDFs for Na$^+$, Rb$^+$, and Sr$^{2+}$ in the 0\% ionized membrane.} We plot the cation-amine RDFs. Coordination shell cutoffs are red dashed lines.}
    \label{SI:fig:membrane_RDFs_0pct_3}
\end{figure}

 \begin{figure}[H]
    \centering
    \includegraphics[width=\textwidth]{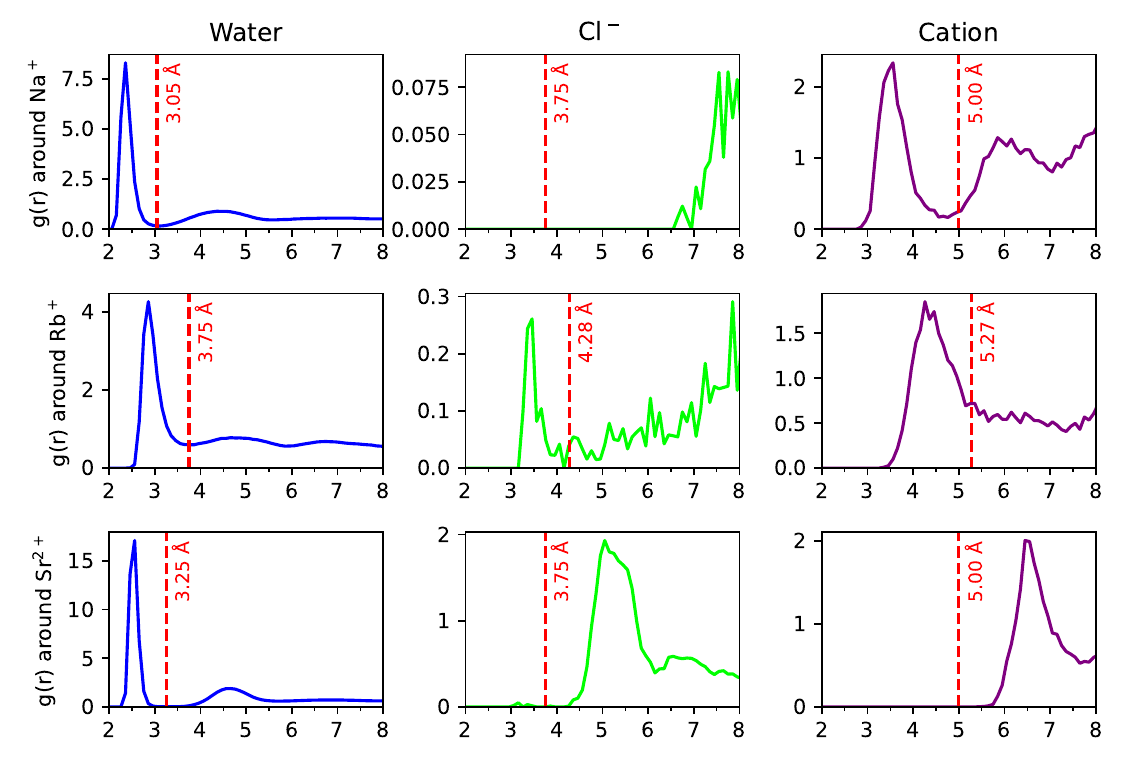}
    \caption{\textbf{RDFs for Na$^+$, Rb$^+$, and Sr$^{2+}$ in the 25\% ionized membrane.} We plot the cation-water, cation-anion, and cation-cation RDFs. Coordination shell cutoffs are red dashed lines.}
    \label{SI:fig:membrane_RDFs_25pct_1}
\end{figure}

 \begin{figure}[H]
    \centering
    \includegraphics[width=\textwidth]{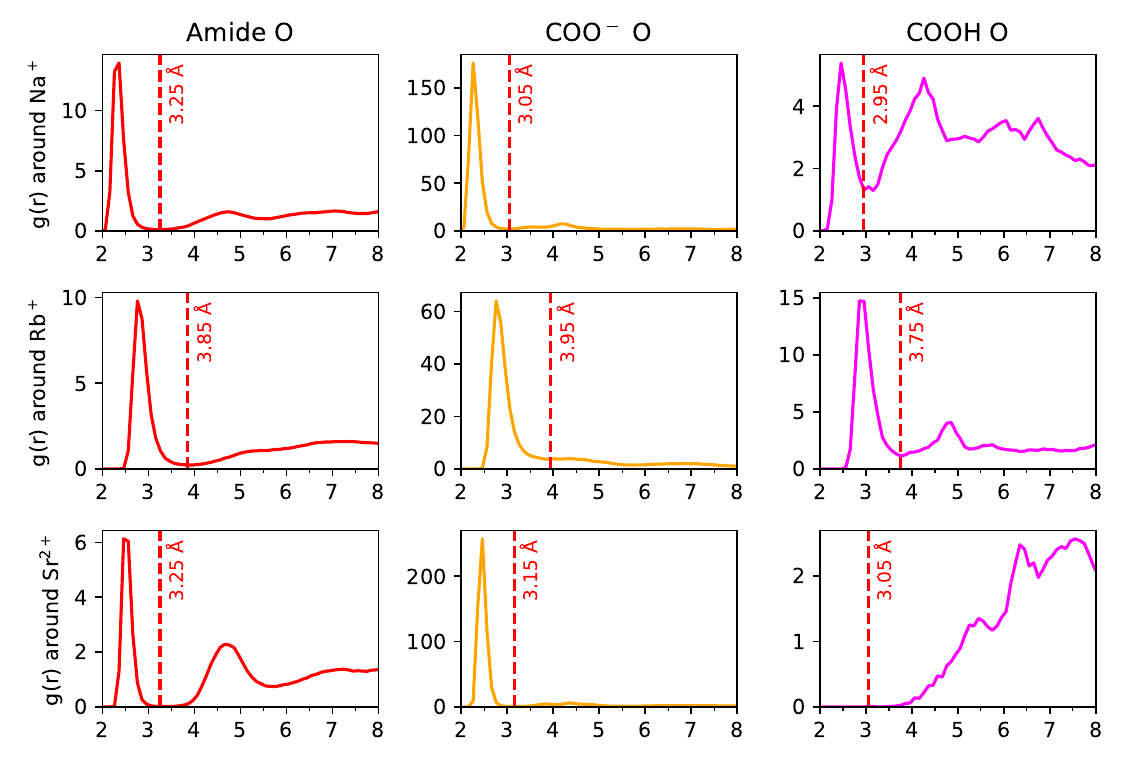}
    \caption{\textbf{RDFs for Na$^+$, Rb$^+$, and Sr$^{2+}$ in the 25\% ionized membrane.} We plot the cation-amide oxygen, cation-carboxylate oxygen, and cation-carboxylic acid oxygen RDFs. Coordination shell cutoffs are red dashed lines.}
    \label{SI:fig:membrane_RDFs_25pct_2}
\end{figure}

 \begin{figure}[H]
    \centering
    \includegraphics[width=0.333\textwidth]{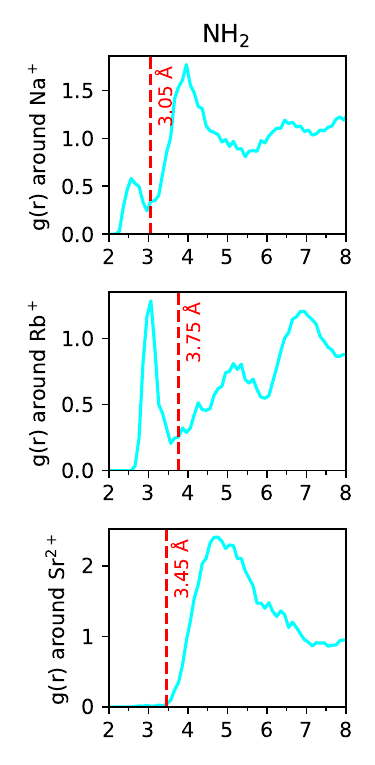}
    \caption{\textbf{RDFs for Na$^+$, Rb$^+$, and Sr$^{2+}$ in the 25\% ionized membrane.} We plot the cation-amine RDFs. Coordination shell cutoffs are red dashed lines.}
    \label{SI:fig:membrane_RDFs_25pct_3}
\end{figure}

 \begin{figure}[H]
    \centering
    \includegraphics[width=\textwidth]{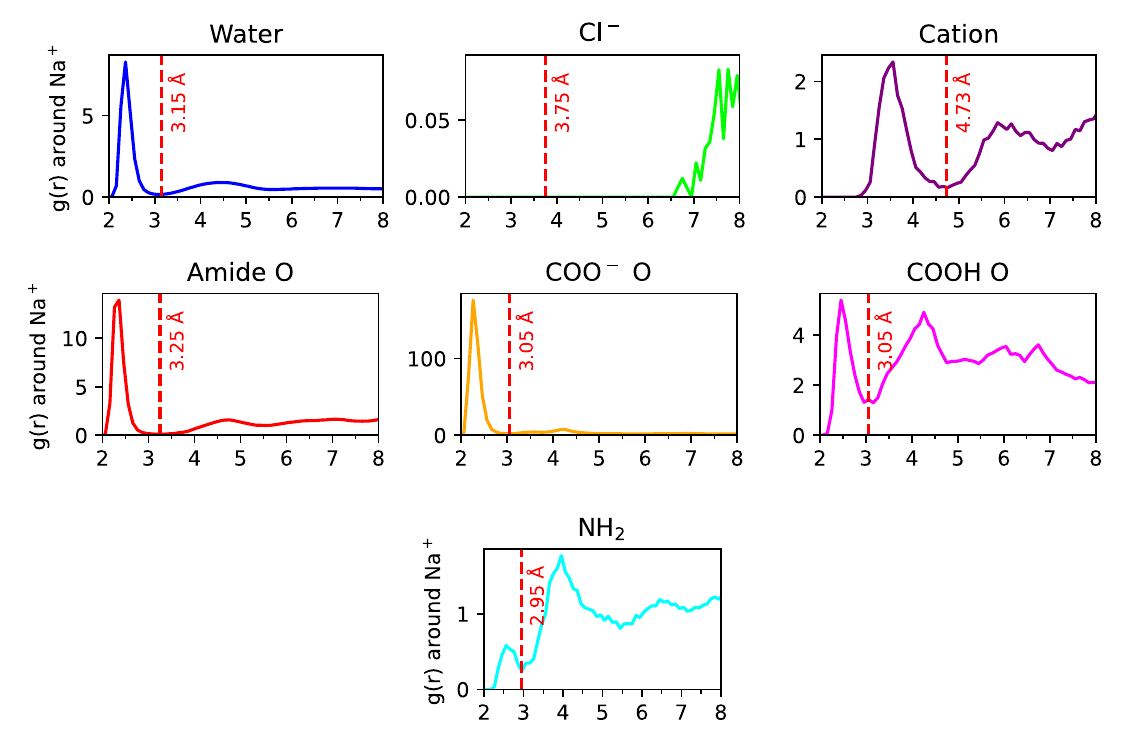}
    \caption{\textbf{RDFs for Na$^+$ in the 50\% ionized membrane hydrated with TIP3P water.} We plot the cation-water, cation-anion, cation-cation, cation-amide oxygen, cation-carboxylate oxygen, cation-carboxylic acid oxygen, and cation-amine RDFs. Coordination shell cutoffs are red dashed lines.}
    \label{SI:fig:membrane_RDFs_TIP3P}
\end{figure}

\section{Example snapshots of cation-cation coordination}

 \begin{figure}[H]
    \centering
    \includegraphics[width=0.75\textwidth]{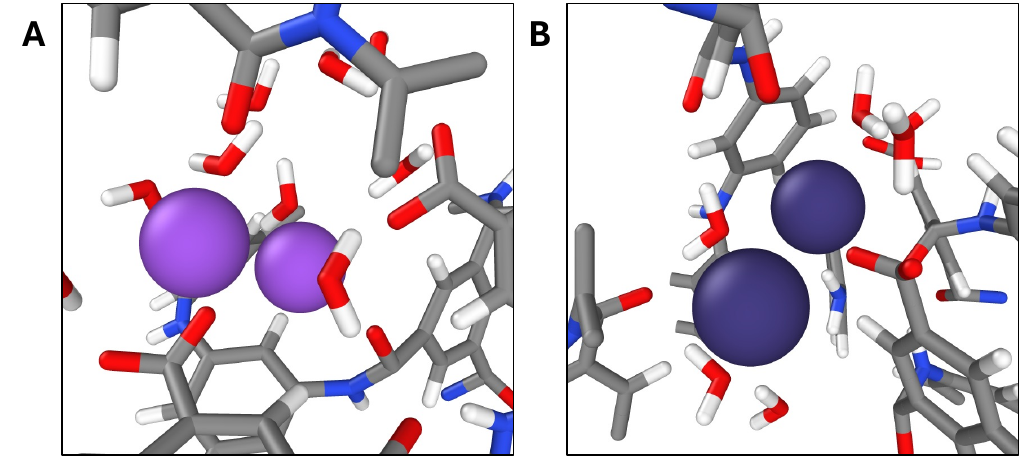}
    \caption{\textbf{Snapshots of cation-cation coordination in the membrane.} (\textbf{A}) We show an example of Na$^+$-Na$^+$ coordination. (\textbf{B}) We show an example of Rb$^+$-Rb$^+$ coordination. In both these snapshots, the cations coordinated tightly with close carboxylate oxygens.}
    \label{SI:fig:cation-cation_coordination}
\end{figure}

\bibliography{local_environment}

\end{document}